\documentclass[review,3p,number,sort&compress, times]{elsarticle}

\usepackage{amssymb}
\usepackage{color}
\usepackage{epstopdf}
\usepackage{booktabs}
\usepackage{paralist}

\usepackage{amsthm}
\usepackage{amsmath}
\usepackage{float}
\usepackage{longtable}
\usepackage[usenames,dvipsnames]{xcolor}

\usepackage{mdframed} 
\usepackage{nomencl} 
\usepackage{multicol} 
\usepackage{titlesec}
\usepackage{xpatch}
\usepackage{nomencl} 
\makenomenclature

\renewcommand*\nompreamble{\begin{multicols}{2}}
\renewcommand*\nompostamble{\end{multicols}}
 
\patchcmd{\thenomenclature}
\makenomenclature
\usepackage{etoolbox}
\renewcommand\nomgroup[1]{%
  \item[\bfseries
  \ifstrequal{#1}{A}{Integer Variables}{%
  \ifstrequal{#1}{B}{Binary Variables}{%
  \ifstrequal{#1}{C}{Real-Valued Parameters}{%
  \ifstrequal{#1}{D}{Real-Valued Variables}{
  \ifstrequal{#1}{E}{Energy-Water Variables}{}}}}}%
]}
\theoremstyle{definition}

\titleformat{\paragraph}
{\normalfont\normalsize\bfseries}{\theparagraph}{1em}{}
\titlespacing*{\paragraph}
{0pt}{3.25ex plus 1ex minus .2ex}{1.5ex plus .2ex}
\journal{Applied Energy}

\begin{document}
\begin{frontmatter}
\title{An Enterprise Control Methodology for the Techno-Economic Assessment of the Energy Water Nexus}
\author[1]{Steffi O. Muhanji\corref{cor1}\fnref{fn1}\corref{cor1}}\ead{steffi.o.muhanji.th@dartmouth.edu}
\author[2]{Amro M. Farid\fnref{fn2}} \ead{amro.m.farid@dartmouth.edu}
\fntext[fn1]{Graduate Student at the Thayer School of Engineering at Dartmouth}
\fntext[fn2]{Associate Professor at the Thayer School of Engineering at Dartmouth}
\address{Thayer School of Engineering, 14 Engineering Dr., Hanover, NH 03755}

\begin{abstract}
In recent years, the energy-water nexus literature has recognized that the electricity and water infrastructure that enable the production, distribution, and consumption of these two precious commodities is fundamentally intertwined.  Electric power is used to produce, treat, distribute, and recycle water while water is used to generate and consume electricity.  In the meantime, significant attention has been given to renewable energy integration within the context of global climate change.   While these two issues may seem unrelated, their resolution is potentially synergistic in that renewable energy technologies not only present low CO$_2$ emissions but also low water-intensities.  Furthermore, because water is readily stored, it has the potential to act as a flexible energy resource on both the supply as well as the demand-side of the electricity grid.  Despite these synergies, the renewable energy integration and energy-water nexus literature have yet to methodologically converge to systematically address potential synergies.  This paper advances an enterprise control methodology as a means of assessing the techno-economic performance of the energy water nexus.  The enterprise control methodology has been developed in recent years to advance the methodological state of the art of renewable energy integration studies and used recently to carry out a full-scale study for the Independent System Operator (ISO) New England system.  The methodology quantifies day-ahead and real-time energy market production costs, dispatched energy mixes, required operating reserves, levels of curtailment, and grid imbalances.  This energy-water nexus methodological extension now includes flexible water-energy resources within the grid's energy resource portfolio and quantifies the amounts of water withdrawn and consumed.   The simulation methodology is demonstrated on a modified version of the RTS-96 (RTS-GMLC) test case.  
\end{abstract}


\end{frontmatter}


\printnomenclature
%
\nomenclature[A]{$k$, $n$}{generator index, bus index}
\nomenclature[A]{$l$, $t$}{line index, time index}
\nomenclature[A]{$m$}{active demand response (DR) index}
\nomenclature[A]{$x$}{penalty variable index}
\nomenclature[A]{$\mathcal{L}$,$\mathcal{V}$}{load index,solar index}
\nomenclature[A]{$\mathcal{H}$,$\mathcal{W}$}{hydro index,wind index}
\nomenclature[A]{$N_{G},N_{D}$}{number of generators, active DR}
\nomenclature[A]{$N_{X}, N_{B}$}{number of penalty variables and buses}
\nomenclature[A]{$N_{\mathcal{L}},N_{\mathcal{W}}$}{number of loads, and wind units}
\nomenclature[A]{$N_{\mathcal{V}},N_{\mathcal{H}}$}{number of solar, and hydro units}
\nomenclature[A]{$A_{n(k,s,x,l,\mathcal{L},\mathcal{W},\mathcal{V},\mathcal{H})}$}{bus incidence matrices}
\nomenclature[B]{$w_{Gkt}$}{ON/OFF state of generator $k$ at time $t$}
\nomenclature[B]{$u_{Gkt}$}{start-up state of generator $k$ at time $t$}
\nomenclature[B]{$v_{Gkt}$}{shutdown state of generator $k$ at time $t$}
\nomenclature[B]{$w_{Dmt}$}{ON/OFF state of DR unit $m$ at time $t$}
\nomenclature[B]{$u_{Dmt}$}{start-up state of DR unit $m$ at time $t$}
\nomenclature[B]{$v_{Dmt}$}{shutdown state of DR unit $m$ at time $t$}
\nomenclature[B]{$w_{Pst}$}{discharging state of storage \textit{s} at time \textit{t}}
\nomenclature[B]{$w_{Sst}$}{charging state of storage \textit{s} at time \textit{t}}
\nomenclature[C]{$C_{Uk}$}{startup costs ($\$$) of generator $k$}
\nomenclature[C]{$C_{Dk}$}{shutdown costs ($\$$) of generator $k$}
\nomenclature[C]{$C_{Um}$}{startup costs ($\$$) of DR unit $m$}
\nomenclature[C]{$C_{Dm}$}{shutdown costs ($\$$) of DR unit $m$}
\nomenclature[C]{$C_{Fk}$}{fixed costs ($\frac{\$}{hr}$) of gen $k$}
\nomenclature[C]{$C_{Fm}$}{fixed costs ($\frac{\$}{hr}$) of DR unit $m$}
\nomenclature[C]{$C_{Lk}$}{linear costs ($\frac{\$}{MWhr}$) of gen $k$}
\nomenclature[C]{$C_{Qk}$}{quadratic costs ($\frac{\$}{MW^2hr}$) of gen $k$}
\nomenclature[C]{$C_{Qm}$}{quadratic costs ($\frac{\$}{MW^2hr}$) of DR unit $m$}
\nomenclature[C]{$C_{Lm}$}{linear cost ($\frac{\$}{MWhr}$) of DR unit $m$}
\nomenclature[C]{$C_{Qx}$}{quadratic cost ($\frac{\$}{MW^2hr}$) of penalty factor $x$}
\nomenclature[C]{$C_{\mathcal{L}}$,$C_{\mathcal{W}}$}{curtailment costs ($\frac{\$}{MWhr}$) for load and wind}
\nomenclature[C]{$C_{\mathcal{V}}$,$C_{\mathcal{H}}$}{curtailment costs ($\frac{\$}{MWhr}$) for solar and hydro}
\nomenclature[C]{$d_{\mathcal{L}}$}{curtailable fraction for load}
\nomenclature[C]{$d_{\mathcal{W}},d_{\mathcal{V}}$}{curtailable fraction for wind, and solar}
\nomenclature[C]{$d_{\mathcal{H}}$}{curtailable fraction for hydro}
\nomenclature[C]{$\overline{P_k}, \underline{P_k}$}{power limits of generator $k$ ($MW$)}
\nomenclature[C]{$\overline{R_k}, \underline{R_k}$}{ramping limits of generator $k$ ($MW/min$)}
\nomenclature[C]{$\overline{P_m}, \underline{P_m}$}{power limits of DR unit $m$ ($MW$)}
\nomenclature[C]{$\overline{R_m}, \underline{R_m}$}{ramping limits of DR unit $m$ ($MW/min$)}
\nomenclature[C]{$\overline{P^+_s}, \underline{P^+_s}$}{power limits of storage unit \textit{s} ($MW$)}
\nomenclature[C]{$\overline{P^-_s}, \underline{P^-_s}$}{pumping limits of storage unit \textit{s} ($MW$)}
\nomenclature[C]{$\overline{E_s}, \underline{E_s}$}{energy capacity limits of storage unit \textit{s} ($MWh$)}
\nomenclature[C]{$\overline{R_{\mathcal{W}}}, \underline{R_{\mathcal{W}}}$}{ramping limits of wind unit $\mathcal{W}$ ($MW/min$)}
\nomenclature[C]{$\overline{R_{\mathcal{V}}}, \underline{R_{\mathcal{V}}}$}{ramping rates of solar unit $\mathcal{V}$ ($MW/min$)}
\nomenclature[C]{$\overline{R_{\mathcal{H}}}, \underline{R_{\mathcal{H}}}$}{ramping rates of hydro unit $\mathcal{H}$ ($MW/min$)}
\nomenclature[C]{$B_{nl}$}{incidence matrix of branches to buses}
\nomenclature[C]{$T_h$}{SCUC scheduling time step (normally, $1h$)}
\nomenclature[C]{$T_m$}{SCED time step (normally, $5-10mins$)}
\nomenclature[C]{$P_{res}$}{load-following reserve requirements ($MW$)}
\nomenclature[C]{$R_{res}$}{ramping reserve requirements ($MW/min$)}
\nomenclature[C]{$\gamma$}{\% transmission losses}
\nomenclature[C]{$\mathbb{C}_F$}{Cost of fuel in $\$/MJ$}
\nomenclature[D]{$\epsilon_{s}$}{reservior level of storage $s$ at $t=0$}
\nomenclature[D]{$P_{kt}$}{power output of generator \textit{k} at time \textit{t}}
\nomenclature[D]{$P_{k}$}{SCED power output of generator \textit{k}}
\nomenclature[D]{$P_{mt}$}{demand response level at time \textit{t} ($MW$)}
\nomenclature[D]{$P_{xt}$}{penalty variable at time \textit{t} ($MW$)}
\nomenclature[D]{$\hat{P}_{\mathcal{L}t}$}{day-ahead load forecast at time \textit{t} ($MW$)}
\nomenclature[D]{$\hat{P}_{\mathcal{W}t}$}{day-ahead wind forecast at time \textit{t} ($MW$)}
\nomenclature[D]{$\hat{P}_{\mathcal{V}t}$}{day-ahead solar forecast at time \textit{t} ($MW$)}
\nomenclature[D]{$\hat{P}_{\mathcal{H}t}$}{day-ahead hydro forecast at time \textit{t} ($MW$)}
\nomenclature[D]{$\tilde{P}_{\mathcal{L}}$}{real-time load forecast ($MW$)}
\nomenclature[D]{$\tilde{P}_{\mathcal{W}}$}{real-time wind forecast ($MW$)}
\nomenclature[D]{$\tilde{P}_{\mathcal{V}}$}{real-time solar forecast ($MW$)}
\nomenclature[D]{$\tilde{P}_{\mathcal{H}}$}{real-time hydro forecast ($MW$)}
\nomenclature[D]{$w_{\mathcal{L}t}$}{percentage of load to curtail at time \textit{t}}
\nomenclature[D]{$w_{\mathcal{W}t}$}{percentage of wind to curtail  at time \textit{t}}
\nomenclature[D]{$w_{\mathcal{V}t}$}{percentage of solar to curtail at time \textit{t}}
\nomenclature[D]{$w_{\mathcal{H}t}$}{percentage of hydro to curtail at time \textit{t}}
\nomenclature[D]{$P^+_{st}$}{discharging level of storage \textit{s} at time \textit{t} ($MW$)}
\nomenclature[D]{$P^-_{st}$}{charging level of storage \textit{s} at time \textit{t} ($MW$)}
\nomenclature[D]{$E_{st}$}{reservoir level of storage $s$ at $t\geq 1$ ($MWh$)}
\nomenclature[D]{$F_{lt}$}{power-flow through branch \textit{l} at time \textit{t}}
\nomenclature[D]{$P_k$}{current  power for the SCED ($MW$)}
\section{Introduction}
Water security is one of the main challenges facing mankind today. Due to the effects of  climate change on hydrology patterns, the amount of available freshwater resources is quickly declining\cite{Rogers:2013:00,Averyt:2013:00}. It is approximated that only $200,000km^3$--1\% of all freshwater is available for human consumption and utilization\cite{Kanyerere:2018:00}. This includes water required for all day-to-day human needs as well as water needed for the agriculture, manufacturing, and electric power sectors\cite{Kanyerere:2018:00,Dodder:2016:00,Al-Ansari:2017:00}. With the expected population growth and industrialization of developing countries, both the energy and water demand per capita are expected to rise significantly\cite{Al-Ansari:2017:00}. As a result of these challenges, being able to efficiently utilize available water resources and prevent over-exploitation is imperative\cite{Kanyerere:2018:00,Dodder:2016:00}. On the one hand, these challenges call for better ways of managing available water resources whether it is in the improvement of water treatment standards, flue gas management, or infrastructure upgrades. On the other hand, better management of water-intensive industries such as the electric power sector would go a long way to minimize their strain on available water resources. Flexible control of the electricity supply system\cite{Al-Nory:2014:00} is particularly crucial within the context of renewable energy integration studies given that renewables 1). are highly variable 2). have very low life-cycle water consumption, and 3). require the grid to have flexible operating capability to be able to respond to variability of supply. The study of the energy-water nexus must, therefore, converge with renewable energy integration studies.  

In recent years, the energy-water nexus literature has recognized that the water and electricity production, distribution, and consumption systems are fundamentally intertwined\cite{Rogers:2013:00,Kanyerere:2018:00,Dodder:2016:00,Al-Ansari:2017:00}. The electricity industry is inherently dependent on the adequate supply of water to support generation whether its in cooling thermal power plants, hydroelectric power generation, or in the extraction of raw fuels such as natural gas\cite{Rogers:2013:00,Kanyerere:2018:00,Dodder:2016:00,Al-Ansari:2017:00}. Thermal power plants withdraw large quantities of water for cooling purposes and depending on the type of cooling technology, a significant amount of this water is lost through \emph{evaporation} or \emph{blowdown}\cite{Averyt:2013:00,Dodder:2016:00}. To illustrate, a recent study estimates that water withdrawals by electricity generating facilities in 2010 constituted 45\% of the overall freshwater withdrawals in the United States with approximately 2\% of that water being consumed as a result\cite{Dodder:2016:00}. In addition to cooling purposes, large quantities of water are utilized in the extraction of raw fuels for electricity generation\cite{Al-Aboosia:2018:00,Armstrong:2018:00}. A recent study reported that the water consumption (in liters per gigajoule -- L/GJ) for worldwide production of carbon-based and nuclear fuels is as follows: 1) traditional oil (3--7 L/GJ); 2) oil from oil sands (70--1800 L/GJ); 3) conventional natural gas (minimal water use); 4) shale gas (36--54 L/GJ); 5) coal (5--70 L/GJ); and 6) uranium (4--22 L/GJ)\cite{Armstrong:2018:00}. Given that in 2015, 76.9\% of the world's total electricity was generated from oil, coal, natural gas and nuclear fuels while 16\% came from hydroelectricity\cite{IEA:2017:01}, reducing the water intensity of these generation processes is crucial to ensuring water security. Similarly, significant electric power is required to support water production and distribution needs such as desalination, waste-water treatment and recycling, and pumping\cite{Rogers:2013:00,Kanyerere:2018:00,Dodder:2016:00,Al-Ansari:2017:00}. With this level of coupling, significant synergies could be realized by studying the two systems holistically.  

In the meantime, significant attention has been given to the integration of renewable energy resources into the electricity grid as a means of decarbonizing the electricity supply system. Due to concerns about climate change, solar and wind installations are steadily increasing while coal, nuclear and oil power plants are slowly being retired\cite{IEA:2017:00}.  Recent studies have shown that variable energy resources (VERs) such as solar and wind possess dynamics that span multiple time scales and hence, affect different layers of power system's control\cite{Curtright:2008:01,Apt:2007:00,Milano:2010:17}. These findings illustrate that the traditional power system's hierarchical control structure is no longer sufficient to ensure the reliability of the system especially as the penetration of VERs continues to grow\cite{Farid:2014:SPG-J26}. Additionally, these studies have also confirmed that due to a high penetration of VERs, operators are forced to rely on manual curtailment of such resources to balance the net load\cite{Wolter:2012:00}. In addition, forecast errors of VERs have been shown to increase infeasible dispatches in the real-time market\cite{Wei:2016:00,Muhanji:2018:SPG-J37}. Another key conclusion of these integration studies is that the intermittency and uncertainty of VERs is likely to increase the reserve requirements and hence the marginal production cost of electricity\cite{DeMeo:2005:00,Smith:2007:00,Bird:2012:00,Brouwer:2014:00}. These factors pose many challenges to grid operators both at the distribution and transmission level. 

While the challenges of renewable generation and energy-water-nexus may seem unrelated, their resolution is potentially synergistic.
Renewable energy technologies not only present low $CO_2$ emissions, but they also have low water-intensities. Furthermore, since water is easily stored, it has the potential to act as a flexible energy resource on both the supply-side as well as the demand-side of the electricity grid\cite{Lubega:2016:00}. As a consequence of decarbonization and low gas prices, a lot of new natural gas power plants are being installed to replace the retired coal and oil generation facilities\cite{IEA:2017:05}. However, natural gas production withdraws and consumes significant amounts of water ($\approx$ $1000m^3$--$30000m^3$ per shale well per year \cite{Al-Aboosia:2018:00}) and hence, cannot be ignored within the context of renewable energy integration \cite{Al-Aboosia:2018:00,Dierauer:2018:00,Al-Douri:2017:00,Kiaghadi:2017:00,Nichols:2017:00,Chang:2015:00,Siddiqi:2013:01}. To meet the required $CO_2$ emission reductions, natural gas production is projected to grow by 44\% between 2011 and 2040 \cite{Sanders:2014:00}. In order to maintain the reliability of the electricity grid with high penetrations of wind and solar, system operators need to flexibly operate generation resources so as to meet the intermittency and uncertainty of solar and wind generation\cite{Farid:2014:SPG-J26}. Additionally, they must have the ability to flexibly control available water-dependent electricity resources and electricity-intensive water processes both to minimize costs and improve the reliability of supply\cite{Farid:2014:SPG-J26}. In this case, water system operators can potentially increase their profits by providing demand-response and ancillary services.    
\subsection{Literature Gap}
Despite the clear synergistic advantages, the renewable energy integration and EWN literature have not yet converged methodologically to systematically address potential synergies. Renewable energy integration studies have focused solely on the operation of electricity markets with large penetrations of VERs\cite{Ela:2009:00,Brouwer:2014:00,Holttinen:2012:01,Holttinen:2013:00,Muzhikyan:2019:SPG-JR04}. A variety of these studies have been case specific and only considered a single unit-commitment/economic dispatch layers of power system control\cite{GE-Energy:2010:01,Shlatz:2011:00,Piwko:2005:00}. Others have taken statistical approaches to determining the forecast errors of wind and solar. A majority of these studies have focused on the acquisition of normal operating reserves such as load-following, regulation, and ramping reserves\cite{Ela:2009:00,Brouwer:2014:00,Holttinen:2012:01,Holttinen:2013:00,Muzhikyan:2019:SPG-JR04}. However, a recent review of integration studies shows major methodological limitations in these studies\cite{Farid:2014:SPG-J26}. First, the quantity of the required reserves is based on the experiences of grid operators which no longer applies to systems with high penetrations of VERs\cite{Holttinen:2008:01,Robitaille:2012:00}. Second, although both the net load variability and forecast error contribute towards normal operating reserves, most studies consider only one of the variables\cite{Ummels:2007:00,Holttinen:2008:01}. Lastly, most studies fail to consider the effects of timescales on the various types of operating reserve quantities.  This same review \cite{Farid:2014:SPG-J26} proposed a holistic approach based on enterprise control to study the full impact of VERs on power system balancing operation and reserve requirements. \emph{Enterprise control} is an integrated and holistic approach that allows operators to improve the technical performance of the grid while realizing cost savings\cite{Farid:2014:SPG-J26}. This approach allows for a multi-timescale analysis of system dynamics and thus, ensures the accurate determination of operating reserves. An application of enterprise control in the form of the Electric Power Enterprise Control System (EPECS) simulator has been proposed in literature and tested on various case studies including the ISO New England system\cite{Muzhikyan:2019:SPG-JR04}.  

In the meantime, the energy-water-nexus literature has come up with individual technologies, policy recommendations and system analysis techniques to study both the electricity and water supply systems. Policy-based studies tend to take a qualitative and sometimes statistical approach while focusing on a specific geographical region\cite{Armstrong:2018:00,Kanyerere:2018:00,Al-Saidi:2018:00,Burnett:2018:00,Dierauer:2018:00,Eren:2018:00,Gurdak:2018:00,Gleick:2017:00,Pincetl:2015:00,Escriva-Bou:2015:00,GEI-Consultants:2012:00,Haynes:1989:01,Mahgary:1993:00,Formiga:2003:00,Diniz:2008:00}. Similarly, system analysis techniques have been case-study driven, geography-specific, rather than generic methodologies that are generally applicable. Some works have studied the water impact of natural gas production, the water-intensity of thermal power plants\cite{Meldrum:2013:00,Averyt:2013:00,Rogers:2013:00,Macknick:2012:00,Macknick:2012:01,Averyt:2011:00}, and the optimization of water pumps\cite{Bagloee:2018:00,Bagirov:2013:00,Ulanicki:2007:00,Lopez-Ibanez:2008:00,Ghelichi:2018:00,Menke:2016:00}. An interesting group of these system analysis techniques are those that co-optimize energy and water resources\cite{Lubega:2014:EWN-J11,Abdulla:2015:EWN-C53,Lubega:2014:EWN-J12,Santhosh:2013:EWN-C16,Santhosh:2012:EWN-C09,Hickman:2017:EWN-J32,Diaz:2017:00,Takahashi:2017:00,Menke:2017:00,Menke:2016:01}. However, the problem with these approaches is that they are single layer optimizations\cite{Santhosh:2013:EWN-C16,Santhosh:2012:EWN-C09,Hickman:2017:EWN-J32}. For example, \cite{Santhosh:2013:EWN-C16} studied only optimal network flow, \cite{Santhosh:2012:EWN-C09} the economic dispatch, and \cite{Hickman:2017:EWN-J32} the unit commitment problem  for a combined water, power, and co-production facilities. Other approaches studied the demand response capabilities of water distribution systems while exploiting key water distribution features such as variable speed pumps to maximize returns and reduce consumption\cite{Diaz:2017:00,Takahashi:2017:00,Menke:2017:00,Menke:2016:01}. Due to a lack of generic techniques, most of these studies are neither generally extensible nor applicable to other case-study geographies. 
\subsection{Original Contribution}
This paper extends the enterprise control approach presented and implemented in \cite{Farid:2014:SPG-J26,Muzhikyan:2015:SPG-C46,Muzhikyan:2015:SPG-C47,Muzhikyan:2015:SPG-J15,Muzhikyan:2015:SPG-J16,Muzhikyan:2015:SPG-J22,Muzhikyan:2016:SPG-J28} so as to assess the techno-economic performance of the energy-water nexus. In recent years, the EPECS methodology has been developed to advance the methodological state of the art of renewable energy integration studies and has been used to carry out a full-scale study in ISO New England\cite{Muzhikyan:2019:SPG-JR04,Muzhikyan:2016:SPG-J28}. The methodology quantifies the dispatched energy mixes, the required operating reserves, levels of curtailment, grid imbalances, and the day-ahead and real-time production costs\cite{Muzhikyan:2019:SPG-JR04,Muzhikyan:2016:SPG-J28}. The energy-water-nexus methodological extension presented in this paper includes flexible water-energy resources within the grid's energy resource portfolio and quantifies the water consumption and withdrawals. For completeness, the methodology presented in this paper is both case and geography independent. The simulation methodology is demonstrated on a modified version of the RTS-96 test case.  
 \begin{figure}[!tbh]
\centering
\includegraphics[scale=0.5]{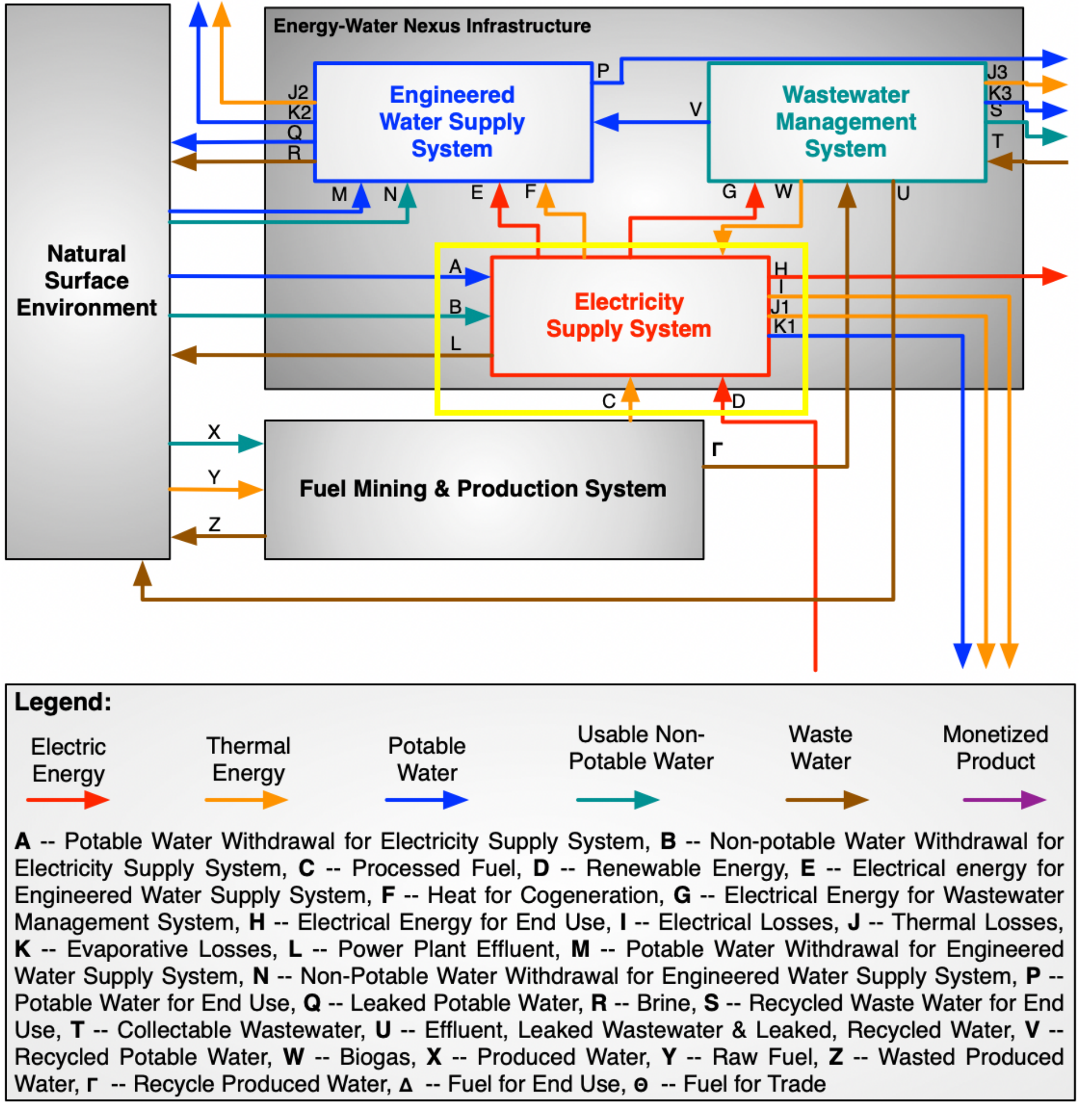}
\caption{This figure shows all the physical flows between the energy water-nexus and the natural surface environment [Adapted from \cite{Abdulla:2015:EWN-C53}].} 
\label{fig:ewnFlows}
\end{figure}
\subsection{Research Scope} This work adopts as its research scope the {\color{blue} yellow} system boundary shown in Fig.~\ref{fig:ewnFlows}.  The traditional electric power system literature does not take into account non-electrical variables at the system boundary.  For example, in power flow analysis, generators are modeled as sources and loads as sinks irrespective of the non-electrical energy flows that they cause upstream or downstream.  In contrast, the system boundary indicated in Fig.~\ref{fig:ewnFlows} explicitly includes all matter and energy flows that enter the electric grid infrastructure.  The energy-water nexus literature, in contrast, often suffers from inconsistencies in the choice of system boundary.  Many of these inconsistencies are caused by the heterogeneity of energy-water resources (or lack thereof) in a methodology tailored to a specific case study geography.  Such studies often fail to recognize that the case study results limit the applicability to other regions and often require that new analytical methodologies be developed as well. Consequently, this study employs a generic methodology that is both \emph{case} and \emph{geography independent} to study the flows in and out of the system boundary especially with a high penetration of VERs. This paper considers the effects of flexible water resources on ensuring the reliability of the electricity grid and on the overall cost of supplying electricity to consumers. The study presents the value of flexible water resources based on how they affect the amount of operating reserves, the total imbalances in the systems, and the electricity market production costs.  In so doing, this work seeks to assess the value of interactions between the electric power system and the natural and built potable water systems.  
\subsection{Paper Outline}
To that end, the rest of this paper is structured as follows: Section \ref{sec:methods} presents the methodological approach for this study. The security-constrained unit commitment (SCUC) and economic dispatch (SCED) formulations are presented in Sections \ref{sec:scuc} and \ref{sec:sced} respectively. The regulation model and the power flow analysis are discussed in Sections \ref{sec:reg} and \ref{sec:water}. A model for studying the water-energy flows is presented in Section \ref{sec:water}. Section \ref{sec:RTS} describes the RTS-GMLC test case and its application to this study. Section \ref{sec:results} presents the results for the case study focusing on operating reserves, balancing performance, fuel consumption and $CO_2$ emissions, water withdrawals and consumption, and the cost implications. Finally, the paper is concluded in Section \ref{sec:conc}.
\section{Methodology}\label{sec:methods}
 \begin{figure*}[!tbh]
\centering
\includegraphics[scale=0.38] {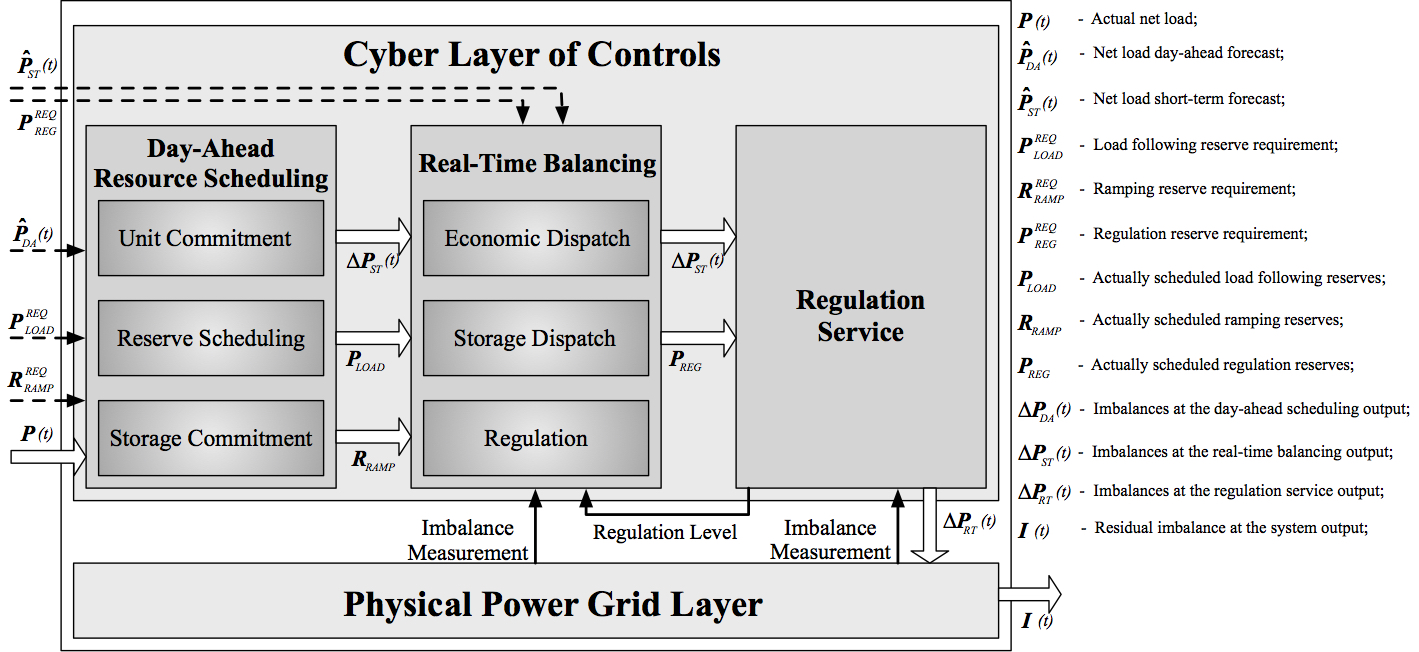}
\caption{The EPECS methodology is used to study the real-time flows for the electricity supply system.} 
\label{fig:epecs}
\end{figure*}
\subsection{Overview}
This paper employs the enterprise control methodology introduced in \cite{Muzhikyan:2015:SPG-J15,Muzhikyan:2015:SPG-J16,Muzhikyan:2015:SPG-J22,Muzhikyan:2016:SPG-J28} as a holistic approach for the techno-economic assessment of newly integrated variable energy resources. The EPECS simulator is a modular simulator that comprises of three control and decision-making layers on top of a physical power grid layer as illustrated in Figure \ref{fig:epecs}. The decision-making and control layers include a resource scheduling layer in the form of a security-constrained unit commitment (SCUC), a balancing layer accomplished through the security constrained economic dispatch (SCED), and a regulation layer. These three layers work together to holistically quantify and address imbalances occurring throughout the electric power system. The enterprise control methodology has been assessed and validated through a set of numerical simulations on various well-known test cases such as the IEEE 11-bus test case, and the IEEE RTS-96 test case\cite{Muzhikyan:2015:SPG-J16}. Most recently, the enterprise control simulator was used to study the impact of various penetrations of wind and solar on the ISO-New England system\cite{Farid:2017:SPG-IP46,Farid:2017:SPG-IP47}. This section presents the enterprise control methodology and extends its application to the techno-economic assessment of the energy-water nexus. The rest of this section is structured as follows: Section~\ref{sec:scuc} defines the SCUC formulation, Section~\ref{sec:sced} presents the SCED formulation, Section~\ref{sec:reg} describes the regulation model and finally Section~\ref{sec:water} presents the mathematics for quantifying the energy-matter flows across the yellow system boundary of Figure~\ref{fig:ewnFlows}.

\subsection{Security-Constrained Unit Commitment (SCUC)}\label{sec:scuc} 
The SCUC commits a set of generators and demand response resources so as to meet the stochastic net load at a minimum cost. It also dispatches storage units, schedules reserves and is executed a day in advance.   The SCUC formulation presented below is adapted from \cite{Muzhikyan:2019:SPG-JR04} in order to accommodate energy and water resources.  

\begin{align}\nonumber
&\min\quad \sum_{t=1}^{24}T_h\bigg(\sum_{k=1}^{N_G}(w_{Gkt}C_{Fk}+C_{Lk}P_{kt}+C_{Qk}P_{kt}^{2}+u_{Gkt}C_{Uk}+v_{Gkt}C_{Dk})+\sum_{s=1}^{N_s}C_{es}E_{st}+(C_{sp}P^+_{st}+C_{sc}P^-_{st})+\ldots&\\\nonumber
&\ldots+\sum_{m=1}^{N_D}(w_{Dmt}C_{Fm}+C_{Lm}P_{mt}+C_{Qm}P_{mt}^2+u_{Dmt}C_{Um}+v_{Dmt}C_{Dm})+\sum_{\mathcal{L}=1}^{N_\mathcal{L}}C_{\mathcal{L}}(1-w_{\mathcal{L}t}d_{\mathcal{L}})\hat{P}_{\mathcal{L}t}+\sum_{x=1}^{N_x}C_{Qx}P_{xt}^2+\ldots&\\\label{objfun:scuc}
&\ldots+\sum_{\mathcal{W}=1}^{N_\mathcal{W}}C_{\mathcal{W}}(1-w_{\mathcal{W}t}d_{\mathcal{W}})\hat{P}_{\mathcal{W}t}+\sum_{\mathcal{V}=1}^{N_\mathcal{V}}C_{\mathcal{V}}(1-w_{\mathcal{V}t}d_{\mathcal{V}})\hat{P}_{\mathcal{V}t}
+\sum_{\mathcal{H}=1}^{N_\mathcal{H}}C_{\mathcal{H}}(1-w_{\mathcal{H}t}d_{\mathcal{H}})\hat{P}_{\mathcal{H}t}\bigg)&\\\nonumber
\end{align}
The objective function in Equation \ref{objfun:scuc} represents the production costs of $N_G$ dispatchable generators, the utility of the $N_D$ demand response resources, the cost of $N_S$ storage resources, the virtual generation cost of $N_\mathcal{L}$ virtual power plants, and the curtailment costs of $N_\mathcal{W}$ wind plants, $N_\mathcal{V}$ solar PV plants, $N_\mathcal{H}$ run-of-river hydro plants.  The objective function also includes a quadratic penalty term $P_{xt}$ that implements a soft constraint in the nodal power balance in each node $x$ on the network. The SCUC objective minimizes the total cost, in dollars, of meeting demand over a period of 24 hours.  
\begin{align}\nonumber
&\sum_{k=1}^{N_G}A_{nk}P_{kt}+\sum_{m=1}^{N_D}A_{nm}P_{mt}-(1+\gamma)\sum_{\mathcal{L}=1}^{N_\mathcal{L}}A_{n\mathcal{L}}(1-w_{\mathcal{L}t}d_{\mathcal{L}})\hat{P}_{\mathcal{L}t}+(1+\gamma)\sum_{\mathcal{W}=1}^{N_\mathcal{W}}A_{n\mathcal{W}}(1-w_{\mathcal{W}t}d_{\mathcal{W}})\hat{P}_{\mathcal{W}t}+\ldots&\\ \nonumber 
&\ldots +\sum_{s=1}^{N_s}A_{ns}(P^+_{st}-P^-_{st})+(1+\gamma)\sum_{\mathcal{V}=1}^{N_\mathcal{V}}A_{n\mathcal{V}}(1-w_{\mathcal{V}t}d_{\mathcal{V}})\hat{P}_{\mathcal{V}t}+(1+\gamma)\sum_{\mathcal{H}=1}^{N_\mathcal{H}}A_{n\mathcal{H}}(1-w_{\mathcal{H}t}d_{\mathcal{H}})\hat{P}_{\mathcal{H}t}+\ldots&\\\label{pb:scuc}
&\ldots +\sum_{x=1}^{N_x}A_{nx}(P_{xt}) = \sum_{l=1}^{N_{L}}B_{nl}F_{lt} \quad\quad\quad\quad\quad\quad\quad\quad\quad\quad\quad\quad\quad\quad\quad\quad\quad\quad\quad \forall_{ t}\\ \nonumber
\end{align}
Equation \ref{pb:scuc} maintains the nodal power balance of all generation, storage and demand-side resource injections with the line flows out of each node.
\begin{align}\label{cc:gen}
&w_{Gkt}\underline{P_{k}}\leq P_{kt}\leq w_{Gkt}\overline{P_{k}}&\forall_{k, t}&\\\label{cc:adr}
&w_{Dmt}\underline{P_{m}}\leq P_{mt}\leq w_{Dmt}\overline{P_{m}}&\forall_{m, t}&\\\label{cc:stor}
&w_{Pst}\underline{P^+_{s}}\leq P^+_{st}\leq w_{Pst}\overline{P^+_{s}}&\forall_{s, t}&\\\label{cc:stor1}
&w_{Sst}\underline{P^-_{s}}\leq P^-_{st}\leq w_{Sst}\overline{P^-_{s}}&\forall_{s, t}&
\end{align}
Equations \ref{cc:gen},\ref{cc:adr}, \ref{cc:stor}, and \ref{cc:stor1} represent the power capacity constraints for dispatchable generation, active DR, and storage resources respectively.

\begin{align}\label{cc:stor2}
&\underline{E_{s}}\leq E_{st}\leq \overline{E_{s}}&\forall_{s, t}&
\end{align}
Furthermore, Equation \ref{cc:stor2} represents the energy capacity constraints of the energy storage resources.  
\begin{align}\label{pb:stor3}
&E_{st}=E_{s,t-1}+(\eta_sP^-_{st}-P^+_{st})\cdot T_h&\forall_{s, t}&
\end{align}
Consequently, Equation \ref{pb:stor3} describes the energy storage state equation of these resources.  
\begin{align}\label{init:stor}
&E_{s0} = \epsilon_s&\forall_{s, t=0}&
\end{align}
Equation \ref{init:stor} describes the initial conditions for the energy storage resources.  
\begin{align}\label{lc:gen}
&w_{Gkt-1}+u_{Gkt}-v_{Gkt} = w_{Gkt}&\forall_{k, t}&\\\label{lc:adr}
&w_{Dmt-1}+u_{Dmt}-v_{Dmt} = w_{Dmt}&\forall_{m, t}&
\end{align}
Equations \ref{lc:gen} and \ref{lc:adr} are the logical state equations governing the switching of dispatchable generators and demand-side resources on and off.  
\begin{align}\label{lc:gen1}
&u_{Gkt}+v_{Gkt} \leq 1&\forall_{k, t}&\\\label{lc:adr1}
&u_{Dmt}+v_{Dmt} \leq 1&\forall_{m, t}&
\end{align}
 Equations \ref{lc:gen1} and \ref{lc:adr1} ensure that the dispatchable generators and active demand-side resources cannot startup and shutdown simultaneously.  
 \begin{align}\label{lc:stor1}
&w_{Pst}+w_{Sst}\leq 1&\forall_{s, t}&\\\label{lc:stor2}
&w_{Pst-1}+w_{Sst}  \leq 1&\forall_{s, t}&\\\label{lc:stor3}
&w_{Pst+1}+w_{Sst}  \leq 1&\forall_{s, t}&
\end{align}
 Equations \ref{lc:stor1}, \ref{lc:stor2}, and \ref{lc:stor3} are charging/discharging rules that constrain the energy storage resources such that they neither charge and discharge simultaneously nor do they switch between charging and discharging without switching off first.  
\begin{align}\label{lc:stor4}
&w_{Ps0}=\omega_{Ps0}&\forall_{s,t=0}&\\\label{lc:stor5}
&w_{Ss0}=\omega_{Ss0}&\forall_{s,t=0}&
\end{align}
Equations \ref{lc:stor4} and \ref{lc:stor5} are the initial conditions of the logical states of the energy storage resources.  

\begin{align}\label{rmp:gen}
&\underline{R_{k}} - \frac{\overline{P_k}}{T_{h}}v_{Gkt}\leq \frac{P_{kt}-P_{k,t-1}}{T_h} \leq \overline{R_{k}}+\frac{\overline{P_k}}{T_h}u_{Gkt}&\forall_{k, t}&\\\label{rmp:adr}
&\underline{R_{m}} - \frac{\overline{P_m}}{T_{m}}v_{Dmt}\leq \frac{P_{mt}-P_{m,t-1}}{T_m} \leq \overline{R_{m}}+\frac{\overline{P_m}}{T_m}u_{Dmt}&\forall_{m,t}&\\\label{rmp:wind}
&\underline{R_{\mathcal{W}}}\leq\frac{(1-w_{\mathcal{W}t}d_{\mathcal{W}})\hat{P}_{\mathcal{W}t}-(1-w_{\mathcal{W},t-1}d_{\mathcal{W}})\hat{P}_{\mathcal{W},t-1}}{T_h} \leq \overline{R_{\mathcal{W}}}&\forall_{\mathcal{W}, t}&\\\label{rmp:solar}
&\underline{R_{\mathcal{V}}}\leq \frac{(1-w_{\mathcal{V}t}d_{\mathcal{V}})\hat{P}_{\mathcal{V}t}-(1-w_{\mathcal{V},t-1}d_{\mathcal{V}})\hat{P}_{\mathcal{V},t-1}}{T_h}\leq \overline{R_{\mathcal{V}}}&\forall_{\mathcal{V}, t}&\\\label{rmp:hydro}
&\underline{R_{\mathcal{H}}}\leq \frac{(1-w_{\mathcal{H}t}d_{\mathcal{H}})\hat{P}_{\mathcal{H}t}-(1-w_{\mathcal{H}t-1}d_{\mathcal{H}})\hat{P}_{\mathcal{H}t-1}}{T_h} \leq \overline{R_{\mathcal{H}}}&\forall_{\mathcal{H}, t}&
\end{align}
Equations \ref{rmp:gen}, \ref{rmp:adr}, \ref{rmp:wind}, \ref{rmp:solar}, and \ref{rmp:hydro} represent the ramping constraints for the dispatchable generators, demand response, wind, solar, and run-of-river hydro resources respectively.  Although wind, solar, and run-of-river hydro resources are variable in nature, they gain a semi-dispatchable nature by virtue of their curtailment capability.   The presence of a curtailment decision implies that such a resource must ramp between two consecutive curtailment values (in time).  This work assumes these variable energy resources can ramp between their maximum and minimum capacities within a single SCED time step of five minutes.  
\begin{align}\nonumber
&\sum_{n=1}^{N_B}\bigg(\sum_{k=1}^{N_G}A_{nk}(w_{kt}\overline{P_k}-P_{kt})+\sum_{\mathcal{W}=1}^{N_\mathcal{W}}A_{n\mathcal{W}}\hat{P}_{\mathcal{W}}w_{\mathcal{W}t}d_{\mathcal{W}}+\sum_{\mathcal{V}=1}^{N_\mathcal{V}}A_{n\mathcal{V}}\hat{P}_{\mathcal{V}}w_{\mathcal{V}t}d_{\mathcal{V}}+\sum_{\mathcal{H}=1}^{N_\mathcal{H}}A_{n\mathcal{H}}\hat{P}_{\mathcal{H}}w_{\mathcal{H}t}d_{\mathcal{H}}\bigg)\geq P_{res}\quad \quad \forall_{k,\mathcal{W},\mathcal{V},\mathcal{H},t}
&\\\label{lfr:up}
&\sum_{n=1}^{N_B}\bigg(\sum_{k=1}^{N_G}A_{nk}(P_{kt}-w_{kt}\underline{P_k})+\sum_{\mathcal{W}=1}^{N_\mathcal{W}}A_{n\mathcal{W}}\hat{P}_{\mathcal{W}}d_{\mathcal{W}}(1-w_{\mathcal{W}t})+\sum_{\mathcal{V}=1}^{N_\mathcal{V}}A_{n\mathcal{V}}\hat{P}_{\mathcal{V}}d_{\mathcal{V}}(1-w_{\mathcal{V}t})+\ldots\\\label{lfr:dn}
&\quad\quad\quad\quad\quad\quad\quad\quad\quad\quad\quad\quad\quad\quad\quad\quad\quad\quad\quad\quad\quad\ldots+\sum_{\mathcal{H}=1}^{N_\mathcal{H}}A_{n\mathcal{H}}\tilde{P}_{\mathcal{H}}d_{\mathcal{H}}(1-w_{\mathcal{H}t})\bigg)\geq P_{res}\quad\quad\forall_{k,\mathcal{W},\mathcal{V},\mathcal{H}, t}
\end{align}
Equations \ref{lfr:up} and \ref{lfr:dn} impose requirements on the quantities of upward and downward load-following reserve requirements. Because wind, solar, and run-of-river hydro resources are semi-dispatchable by virtue of their curtailment capability, the amount of power from their maximum and minimum capacity values is included in the accounting of load following reserves.  
\begin{align}\nonumber
&\sum_{n=1}^{N_B}\bigg(\sum_{k=1}^{N_G}A_{nk}\big(w_{kt}\underline{R_{k}}-R_{kt}\big)+\sum_{m=1}^{N_D}A_{nm}\big(w_{mt}\overline{R_{m}}-R_{mt}\big)+\sum_{\mathcal{W}=1}^{N_\mathcal{W}}A_{n\mathcal{W}}\big(w_{\mathcal{W}t}\overline{R_{\mathcal{W}}}-R_{\mathcal{W}t}\big)+\ldots&\\\label{rmp:up}&\ldots+\sum_{\mathcal{V}=1}^{N_\mathcal{V}}A_{n\mathcal{V}}\big(w_{\mathcal{V}t}\overline{R_{\mathcal{V}}}-R_{\mathcal{V}t}\big)+\sum_{\mathcal{H}=1}^{N_\mathcal{H}}A_{n\mathcal{H}}\big(w_{\mathcal{H}t}\overline{R_{\mathcal{H}}}-R_{\mathcal{H}t}\big)\bigg)\geq R_{res}&\forall_{k,\mathcal{W},\mathcal{V},\mathcal{H}, t}&\\\nonumber
&\sum_{n=1}^{N_B}\bigg(\sum_{k=1}^{N_G}A_{nk}\big(R_{kt} - w_{kt}\underline{R_{k}}\big)+\sum_{m=1}^{N_D}A_{nm}\big(R_{mt} - w_{mt}\underline{R_{m}}\big)+\sum_{\mathcal{W}=1}^{N_\mathcal{W}}A_{n\mathcal{W}}\big(R_{\mathcal{W}t}-w_{\mathcal{W}t}\underline{R_{\mathcal{W}}}\big)+\ldots&\\\label{rmp:dn}
&\ldots+\sum_{\mathcal{V}=1}^{N_\mathcal{V}}A_{n\mathcal{V}}\big(R_{\mathcal{V}t}-w_{\mathcal{V}t}\underline{R_{\mathcal{V}}}\big)+\sum_{\mathcal{H}=1}^{N_\mathcal{H}}A_{n\mathcal{H}}\big(R_{\mathcal{H}t}-w_{\mathcal{H}t}\underline{R_{\mathcal{H}}}\big)\bigg)\geq R_{res}&\forall_{k,\mathcal{W},\mathcal{V},\mathcal{H}, t}&
\end{align}
Finally, Equations \ref{rmp:up} and \ref{rmp:dn} are the upward and downward ramping constraints respectively. Similar to load-following constraints, these reserve constraints include contributions from solar, wind and run-of-river hydro resources.
\subsection{SCED}\label{sec:sced} 
This section provides the mathematical formulation for the security constrained economic dispatch (SCED). The SCED runs every 5 minutes to provide new set-points for dispatchable generators, wind, solar, hydro, and active demand-side resources.  Similar to the SCUC, the objective function for SCED includes a quadratic penalty term to account for cases where nodal power balance cannot be achieved with the existing set of energy resources.  The SCED does not commit any new units.  Instead it ramps up and down already committed dispatchable generators and sets new curtailment levels for solar, wind, run-of-river, and demand-side resources. Unlike the SCUC, the SCED does not re-optimize the energy storage setpoints, but rather uses those calculated in the execution of the SCUC.  The SCED formulation minimizes the following objective function:  
\begin{align}\nonumber
&\min \quad  \frac{T_m}{60}\bigg(\sum_{k=1}^{N_G}(C_{Lk}P_{k}+C_{Qk}P_{k}^{2})+\sum_{\mathcal{L}=1}^{N_\mathcal{L}}C_{\mathcal{L}}(1-w_{\mathcal{L}}d_{\mathcal{L}})\tilde{P}_{\mathcal{L}}+\sum_{s=1}^{N_s}C_{sp}P^+_{s}-C_{sc}P^-_{s}+\ldots&\\\nonumber
&\ldots+\sum_{m=1}^{N_D}(C_{Lm}P_{m}+C_{Qm}P_{m}^2)+\sum_{\mathcal{W}=1}^{N_\mathcal{W}}C_{\mathcal{W}}(1-w_{\mathcal{W}}d_{\mathcal{W}})\tilde{P}_{\mathcal{W}}+\ldots&\\\label{obj:sced}&\ldots+\sum_{\mathcal{V}=1}^{N_\mathcal{V}}C_{\mathcal{V}}(1-w_{\mathcal{V}}d_{\mathcal{V}})\tilde{P}_{\mathcal{V}}+\sum_{\mathcal{H}=1}^{N_\mathcal{H}}C_{\mathcal{H}}(1-w_{\mathcal{H}}d_{\mathcal{H}})\tilde{P}_{\mathcal{H}}+\sum_{x=1}^{N_x}C_{Qx}P_{x}^2\bigg)
\end{align}
This objective function is similar to that of the SCUC except that it optimizes over a single time step every $T_m$ minutes and eliminates the energy storage resource terms. The SCED objective is multiplied by a factor of $\frac{T_m}{60}$ to obtain a cost in dollars rather than \$/hr. 
\begin{align}\nonumber
&\sum_{k=1}^{N_G}A_{nk}P_{k}+\sum_{m=1}^{N_D}A_{nm}P_{m}-(1+\gamma)\sum_{\mathcal{L}=1}^{N_\mathcal{L}}A_{n\mathcal{L}}(1-w_{\mathcal{L}}d_{\mathcal{L}})\tilde{P}_{\mathcal{L}}+(1+\gamma)\sum_{\mathcal{H}=1}^{N_\mathcal{H}}A_{n\mathcal{H}}(1-w_{\mathcal{H}}d_{\mathcal{H}})\tilde{P}_{\mathcal{H}}+\ldots&\\\nonumber
&\ldots+\sum_{s=1}^{N_s}A_{ns}(P^+_{s}-P^-_{s})+(1+\gamma)\sum_{\mathcal{W}=1}^{N_\mathcal{W}}A_{n\mathcal{W}}(1-w_{\mathcal{W}}d_{\mathcal{W}})\tilde{P}_{\mathcal{W}}+(1+\gamma)\sum_{\mathcal{V}=1}^{N_\mathcal{V}}A_{n\mathcal{V}}(1-w_\mathcal{V}d_\mathcal{V})\tilde{P}_\mathcal{V}+\ldots&\\\label{pb:sced}  
&\ldots+\sum_{x=1}^{N_x}A_{nx}P_{x}^2 = \sum_{l=1}^{N_{L}}B_{nl}F_{l}\quad\quad\forall_{k, m, x, n, l, s, \mathcal{W},\mathcal{V},\mathcal{H},\mathcal{L}}
\end{align}
Similarly, the nodal-power balance constraint in \ref{pb:sced} is expressed for a single moment in time.
\begin{align}\label{cced:gen}
&w_{k}\underline{P_{k}}\leq P_{k}\leq w_{k}\overline{P_{k}}&\forall_{k}&\\\label{cced:adr}
&\underline{P_{m}}\leq P_{m}\leq \overline{P_{m}}&\forall_{m}&
\end{align}
Equations \ref{cced:gen} and \ref{cced:adr} are the capacity constraints for the dispatchable generators and the active demand response units.
\begin{align}\label{rmped:gen}
&\underline{R_{k}} \leq \frac{P_{k}-P_{k}^0}{T_m} \leq \overline{R_{k}}&\forall_{k}&\\\label{rmped:adr}
&\underline{R_{m}} \leq \frac{P_{m}-P_{m}^0}{T_m} \leq \overline{R_{m}}&\forall_{m}&\\\label{rmped:wind}
&\underline{R_{\mathcal{W}}}\leq\frac{\tilde{P}_{\mathcal{W}}(1-w_{\mathcal{W}}d_{\mathcal{W}})-\tilde{P}_\mathcal{W}^0}{T_m}\leq \overline{R_{\mathcal{W}}}&\forall_{\mathcal{W}}&\\\label{rmped:solar}
&\underline{R_{\mathcal{V}}}\leq\frac{\tilde{P}_\mathcal{V}(1-w_{\mathcal{V}}d_{\mathcal{V}})-\tilde{P}_\mathcal{V}^0}{T_m} \leq \overline{R_{\mathcal{V}}}&\forall_{\mathcal{V}}&\\\label{rmped:hydro}
&\underline{R_{\mathcal{H}}}\leq\frac{\tilde{P}_\mathcal{H}(1-w_{\mathcal{H}}d_{\mathcal{H}})-\tilde{P}_\mathcal{H}^0}{T_m} \leq \overline{R_{\mathcal{H}}}&\forall_{\mathcal{H}}&
\end{align}
Finally, Equations \ref{rmped:gen}, \ref{rmped:adr}, \ref{rmped:wind}, \ref{rmped:solar}, and \ref{rmped:hydro} are the ramping constraints for the dispatchable generators, the active DR, wind, solar, and run-of-river hydro resources respectively.
\subsection{Regulation Reserves}\label{sec:reg}
Regulation reserves are provided by generation units with automatic generation control (AGC) capability.  As described previously in detail\cite{Muzhikyan:2015:SPG-J15}, the EPECS methodology simulates in 1-minute increments. The regulation service generators respond to imbalances by varying their output in the direction opposite to the imbalance until the imbalance is mitigated or the available regulation capacity is exhausted. The EPECS simulator also uses a virtual slack generator that consumes any mismatch between generation and load to make the steady state power flow equations feasible. The power system imbalances are quantified as the output of the slack generator.

\subsection{Model of Physical Energy and Water Flows in the Electricity Supply System}\label{sec:water}
In order apply the enterprise control model described above to the techno-economic assessment of the energy-water nexus, the physical grid model must be extended to quantify the energy and water flows. More specifically, this section provides a methodology by which to calculate the energy and water flows (A through K, and W) that cross the yellow system boundary depicted in Figure \ref{fig:ewnFlows}.  For simplicity, all calculations are done in SI units as indicated in the nomenclature.  
\subsubsection{DC Power Flow Analysis Model}\label{sec:dcpfa}
The heart of the electricity supply system model is a DC power flow analysis model that is solved at each minute time step.   In that regard, for a given minute-time step t, the flow of electric power is assumed to follow Equation \ref{pb:sced} in Section \ref{sec:sced}.  This model couples all of the system's electrical variables  in generation, transmission, and consumption. The remainder of this section, relates the energy and water flows in Figure \ref{fig:ewnFlows} to these electrical variables.

\subsubsection{C: Processed Fuel Used}
One of the main roles of the electricity supply system is to convert processed fuels (C) into electrical energy and deliver it to meet electrical end-uses (E,F,G,H).  The fuel flow rate $C_{k}(t)$ ($kg/min$) for a given dispatchable generator $k$ is extracted from the generator's fuel curve used in the objective function of the SCUC (Equation \ref{obj:sced}).  
\begin{align}\label{ewn:totEnergyin}
 &C_{k}(t) = \frac{C_{Qk}{P}_{k}^2 + C_{Lk}{P}_{k}+w_{Gk}C_{Fk}}{60\mathbb{C}_FD_{fk}}&\forall{k,t}\\\label{ewn:CO2}
 &\kappa_{k}(t) = C_{k}(t) \times D_{fk}\times\xi_{f}&\forall{k,t}
\end{align}
where ${P}_{k}$ is the real-time power generation of generator $k$, $\mathbb{C}_F$ is the fuel cost in $\$/MJ$, $D_{fk}$ ($MJ/kg$) is the fuel energy density and 60 is the conversion from hours to minutes. From the fuel consumed, the $CO_2$ emissions can be calculated as shown in equation~\ref{ewn:CO2}. Whereby $\kappa_{k}(t)$ is the $CO_2$ emitted by generator $k$ in $kg/min$ and $\xi_{f}$ is the $CO_2$ emissions constant for fuel $f$ in $kg/MJ$.
\subsubsection{D: Renewable Energy Delivered}\label{sec:ver}
In the EPECS simulator, the real-time solar PV $P_{\mathcal{V}}(t)$, run-of-river hydro $P_{\mathcal{H}}(t)$, and wind generation $P_{\mathcal{W}}(t)$ are exogeneous quantities drawn from input temporal profile data.  This data is scaled by varying five parameters:  \emph{penetration level} ($\pi$), \emph{capacity factor} ($\gamma$), \emph{variability} ($A$), \emph{day- ahead forecast error} ($\hat{\epsilon}$) and \emph{short-term ($\tilde{\epsilon}$) forecast error}. The penetration level and capacity are used to determine the actual output of the variable energy resource (VER). The VER output is normalized to a unit capacity factor. The day ahead (mean absolute) forecast error with a 1-hour resolution is used in the SCUC formulation while the short-term (mean absolute) forecast error with a 5-minute resolution is used in the SCED formulation.  The interested reader is referred to earlier works for further details\cite{Muzhikyan:2015:SPG-J15,Muzhikyan:2016:SPG-J28}. Both solar and wind can be curtailed in order to balance the grid in the real-time. In this work, the renewable energy delivered (D) (to the electric grid) is the sum of the curtailed wind and solar generation as \emph{endogeneous} results of the SCED (\ref{sec:sced}) and SCUC (\ref{sec:scuc}) models. 
\begin{align}\label{ewn:ren}
 &D(t) = \sum_{\mathcal{V}=1}^{N_{\mathcal{V}}}(1-w_{\mathcal{V}t}d_\mathcal{V})P_{\mathcal{V}}(t)+\sum_{\mathcal{W}=1}^{N_{\mathcal{W}}}(1-w_{\mathcal{W}t}d_\mathcal{W})P_{\mathcal{W}}(t)&
\end{align}
Equation~\ref{ewn:ren} represents the total renewable energy delivered in ($MW$) at each minute time-step $t$.  
\subsubsection{E: Electrical Energy for Water Supply System}\label{ewn:portelec}
The electrical energy consumed by the water supply system (E) is a fraction $d_{\mathcal{L}w}$ of the total electricity demand, and is consequently an exogeneous quantity drawn from input temporal profile data.  This portion of the demand acts as a virtual power plant and can be incentivized downwards as part of the demand response scheme integrated into the SCUC and SCED models above.  
\begin{align}\label{ewn:potH2O}
 &E(t) = \sum_{\mathcal{L}w=1}^{N_{\mathcal{L}w}}(1-w_{\mathcal{L}wt}d_{\mathcal{L}w})P_{\mathcal{L}w}(t)&
\end{align}
The final electrical energy consumed by the water supply system in $MW$ is shown in Equation~\ref{ewn:potH2O} as the uncurtailed amount of water supply electricity demand.
\subsubsection{F: Thermal Energy for Water Desalination}
No multi-stage flash desalination units were included in this study. The reader is referred to several works that have treated this subject in detail\cite{Khawaji:2008:00,Al-Marafie:1989:00,Shakouri:2010:00,Mezher:2011:01,Shakib:2012:00,Ghobeity:2014:00,Al-Aboosia:2018:00,Cardona:2002:00,Pechtl:2003:00,Cardona:2004:00,Sommariva:2007:00,Gebel:2008:00,Khawaji:2008:00,El-Nashar:2008:00}.
\subsubsection{G: Electrical Energy for Wastewater Management System}
Similar to flow E, the electrical energy delivered to the wastewater management system (G) is an exogeneous input to the EPECS simulator and can be incentivized downwards as part of the demand response scheme integrated into the SCUC and SCED models above.  
\begin{align}\label{ewn:npotH2O}
 &G(t) = \sum_{\mathcal{L}ww=1}^{N_{\mathcal{L}ww}}(1-w_{\mathcal{L}wwt}d_{\mathcal{L}ww})P_{\mathcal{L}ww}(t)&
\end{align}
The final electrical energy consumed by the wastewater management system is shown in Equation~\ref{ewn:npotH2O} as the uncurtailed electricity demanded in $MW$ by wastewater management systems.
\subsubsection{H: Electrical Energy for End Use}
The electrical energy delivered for end use (H) is calculated as the total demand minus the electrical demand for the water supply and wastewater management systems as shown in Equation~\ref{ewn:totDem}.
\begin{align}\label{ewn:totDem}
 &H(t) = \sum_{\mathcal{L}=1}^{N_{\mathcal{L}}}P_{\mathcal{L}}(t)-G(t)-E(t)&
 \end{align}

\subsubsection{I: Electrical Losses}
The DC power flow analysis model described in Section \ref{sec:dcpfa} assumes zero electrical losses.  
\subsubsection{J: Thermal Losses}
The thermal losses $J_{k}$ in (MJ/min) of a given power plant $k$ shown in Equation~\ref{ewn:Qtherm} includes all the heat lost to cooling and flue gases. It is calculated from the difference in the net input fuel and the electrical energy generated. A factor of 60 is multiplied by $P_k(t)$ to convert it from MW to MJ/min.
\begin{figure}[!tbh]
\centering
\includegraphics[scale=0.75] {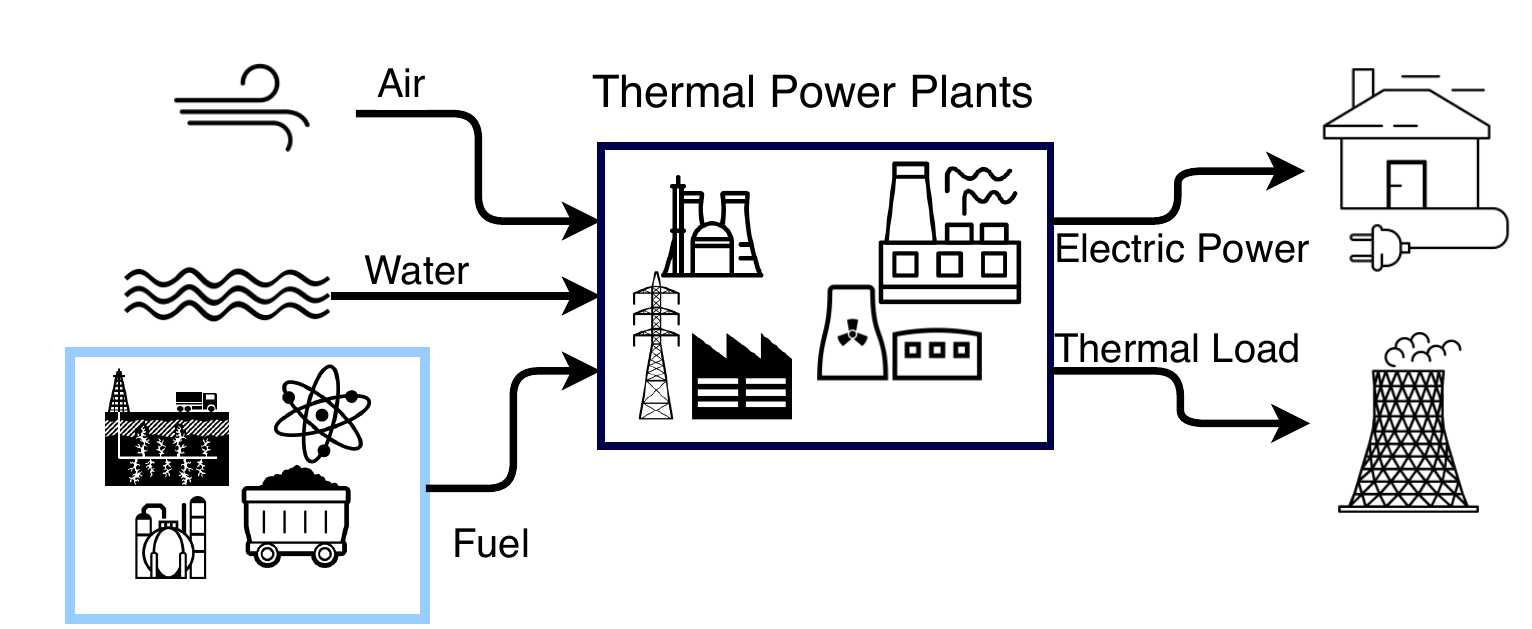}
\caption{Input/Output model for thermal power plants.} 
\label{fig:thermGens}
\end{figure}
\begin{align}\label{ewn:Qtherm}
&J_{k}(t) = C_{k}(t)D_{fk}-60P_{k}(t)&
\end{align}
\subsubsection{A: Portable Water Withdrawal for Electricity Supply System}
\begin{figure}[!t]
\centering
\includegraphics[width=5.2in]{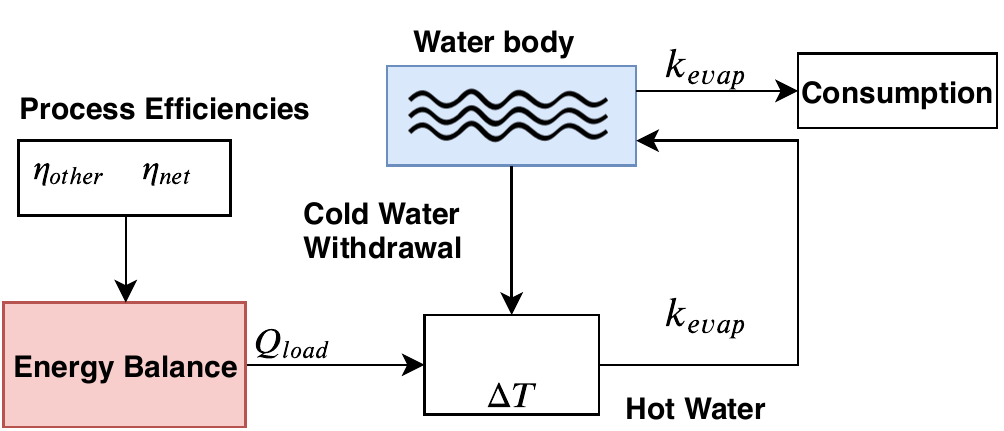}
\caption{Once-through cooling system.} 
\label{fig:onceThrough}
\end{figure}

To study the portable water withdrawal for thermal power plants, this work adopts the system-level generic model (S-GEM) introduced in \cite{Rutberg:2011:00}. The S-GEM was developed to study the water use of fossil fuel, nuclear, geothermal and solar thermal power plants using either steam or combined cycle technologies. The S-GEM model captures the essential physics of cooling processes while minimizing the number of required input parameters and computational complexity. The model is also geography and case-independent; making it ideal for application in this work.  Three main cooling processes are applied in this paper: once-through cooling, wet tower cooling and dry-air cooling.
\paragraph{Once-Through Cooling Systems}
Figure \ref{fig:onceThrough} represents a once-through cooling system. Once-through cooling, also known as open-loop cooling, draws cool water from a water body, passes it through a heat exchanger to cool the thermal load $Q_{Tk}(t)$ (MJ/min) and returns the warm water back to the same water body\cite{Rutberg:2012:00,Integrated-Pollution-Prevention-and-Control-IPPC:2001:00,ECOFYS:2014:00,Eurelectric:1999:00,Tsou:2013:00,Pan:2018:01}. Although this system is simple and has a relatively low cost, it withdraws large quantities of water from the surface environment which may endanger acquatic life through entrainment\cite{Rutberg:2012:00,Integrated-Pollution-Prevention-and-Control-IPPC:2001:00,ECOFYS:2014:00,Eurelectric:1999:00,Tsou:2013:00,Pan:2018:01}.  It also discharges waste-heat and anti-corrosion, scaling, and bio-fouling chemicals back to the water source\cite{Integrated-Pollution-Prevention-and-Control-IPPC:2001:00,ECOFYS:2014:00,Eurelectric:1999:00}.  Evaporative losses through these cooling systems are often negligible and are, therefore, assumed to be zero.  Due to their potentially harmful ecological impacts, once-through cooling systems are less popular for newer generation plants. That said, a lot of older coal, oil and nuclear plants generation still use once-through cooling to dissipate their waste heat. 
\begin{align}\label{ewm:ThermCool}
&Q_{Tk}(t) = J_{k}(t)(1-\eta_{k,other})&\\\label{ewn:otmassflow}
&M_w(t) = \frac{Q_{Tk}(t)}{c_{p,w}\Delta T_{cond}} = J_{k}(t)(1-\eta_{k,other})\frac{1}{c_{p,w}\Delta T_{cond}}&
\end{align}
Equation~\ref{ewm:ThermCool} represents the thermal load $Q_{Tk}(t)$ for a generator $k$ at time $t$ that requires cooling, where $\eta_{k,other}$ represents the fraction of the thermal load $J_{k}(t)$ that is lost through other means (e.g. flue gases). Equation~\ref{ewn:otmassflow} shows the mass flow rate of water in $kg/min$ from the water body where $c_{p,w}$ is the specific heat capacity of water in $MJ/kg\cdot K$ while $\Delta T_{cond}$ is the temperature difference between the cooling water and the process hot water. 
\paragraph{Recirculating Wet Tower Cooling Systems}
 \begin{figure}[!tbh]
\centering
\includegraphics[width=5.2in] {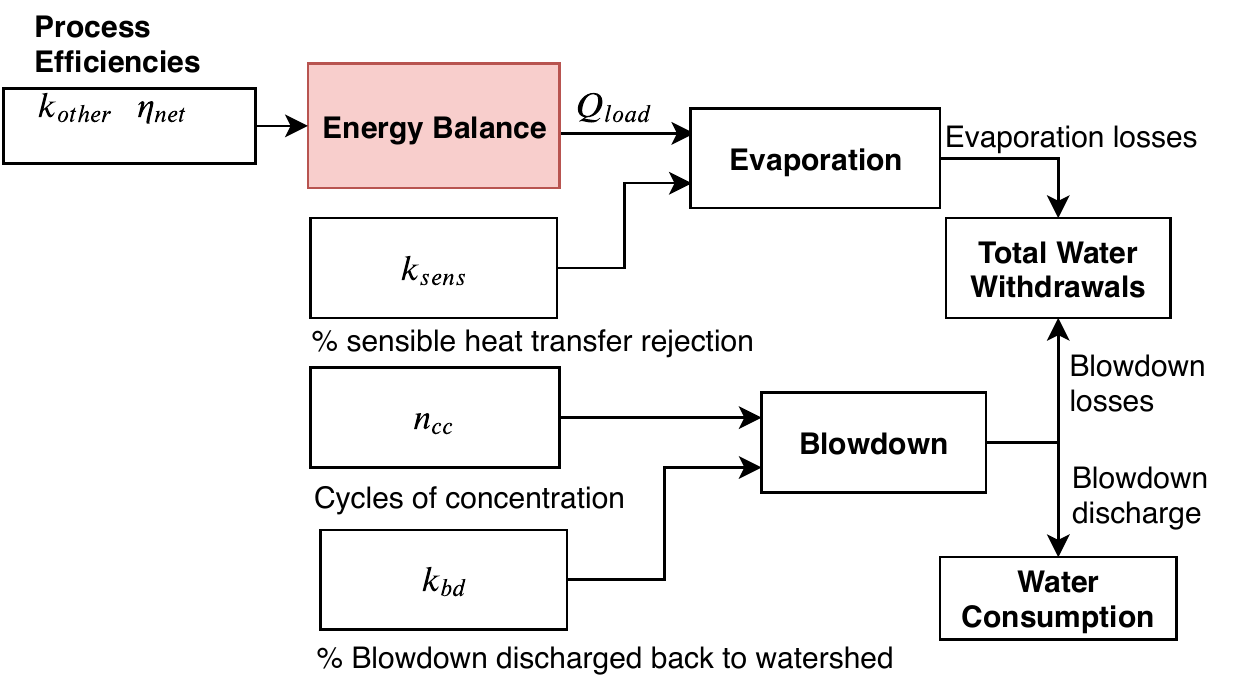}
\caption{Wet tower cooling also known as a recirculating cooling system.} 
\label{fig:recirc}
\end{figure}
Figure \ref{fig:recirc} depicts the flows of a recirculating wet tower cooling system. A recirculating loop of cooling water is sent through the system\cite{Rutberg:2012:00,Integrated-Pollution-Prevention-and-Control-IPPC:2001:00,ECOFYS:2014:00,Eurelectric:1999:00,Tsou:2013:00,Pan:2018:01}.  After cooling water passes through the waste heat exchanger, the now warm water is sprayed down through a lattice-like fill material which increases the surface area through which the water must flow down in the cooling tower\cite{Rutberg:2012:00,Integrated-Pollution-Prevention-and-Control-IPPC:2001:00,ECOFYS:2014:00,Eurelectric:1999:00,Tsou:2013:00,Pan:2018:01}. As the warm water is sprayed down through the fill, a fan or natural draft draws in air from the bottom of the tower up through the fill and out to the environment\cite{Rutberg:2012:00,Integrated-Pollution-Prevention-and-Control-IPPC:2001:00,ECOFYS:2014:00,Eurelectric:1999:00,Tsou:2013:00,Pan:2018:01}. The water and air flow through the tower serves as a heat exchanger to cool the water down before it is recirculated back in the system. 

The bulk of the heat is lost through convective heat transfer from the hot water to the air. $k_{sens}$ represents the fraction of heat lost through sensible heat transfer between air and water\cite{Rutberg:2012:00,Rutberg:2011:00,King:2014:01}. It largely depends on the temperature of the incoming air and less so on other factors such as humidity and atmospheric pressure\cite{Rutberg:2012:00,King:2014:01}. In addition to sensible heat transfer, some of the water evaporates and the latent heat of this evaporation process results in further cooling. A bulk of water consumption in a recirculating cooling system is mainly due to evaporation from the cooling tower \cite{Rutberg:2012:00,Rutberg:2011:00,Integrated-Pollution-Prevention-and-Control-IPPC:2001:00,ECOFYS:2014:00,Eurelectric:1999:00,Tsou:2013:00}.

Additionally, a small percentage of blowdown water is occasionally flushed out of the system to avoid any build up of contaminants. The blowdown may be consumed through evaporation or treated and sent back to the natural surface water system. This study assumes that the entire blowdown is treated and sent back to the natural environment.  Recirculating systems do not withdraw nearly as much water as once-through systems. However, a significant amount of water is consumed through evaporation.  

Given the recirculating nature of this type of cooling system, the total water withdrawal for a recirculating system is assumed to equal the amount of water lost through evaporation and blowdown.
Figure \ref{fig:recirc} best illustrates the process flows for recirculating systems. 
\begin{align}\label{ewn:evaploss}
&\dot{M}_{evap,k}(t) = Q_{Tk}(t)\frac{(1-k_{sens,t})}{h_{fg}}&
\end{align}
The rate of water loss, $kg/min$, through evaporation is computed as shown in Equation~\ref{ewn:evaploss} where $k_{sens}$ is the energy fraction transferred from the hot water to the cool air while $h_{fg}$ is the latent heat of vaporization in units of $MJ/kg$.
\begin{align}\label{ewn:blowdn}
&\dot{M}_{bd,k}(t) = \dot{M}_{evap,k}(t)\bigg(\frac{1}{n_{cc}-1}\bigg)&
\end{align}
The rate of blowdown is represented by Equation~\ref{ewn:blowdn}. Note that the blowdown rate is related to the rate of evaporation $\dot{M}_{evap}$ and the cycles of concentration $n_{cc}$. $n_{cc}$ is a parameter that describes the concentration of impurities in the water circulating through the cooling system relative to that of the makeup water. Typical $n_{cc}$ values used for North American systems range between 2 and 10 cycles of concentration. In this study, an average $n_{cc}$ value of 6 was used. 
\begin{align}\label{ewn:withdrawrecirc}
& \dot{M}_{w,recirc}(t) =  Q_{Tk}(t)\bigg(\frac{1-k_{sens}}{h_{fg}}\bigg)\bigg(1+\frac{1}{n_{cc}-1}\bigg)&
\end{align}
The rate of water lost in $kg/min$ from the cooling tower can be found by combining equations \ref{ewn:Qtherm}, \ref{ewn:evaploss}, and \ref{ewn:blowdn} as shown in Equation~\ref{ewn:withdrawrecirc}.

\paragraph{Dry Air Cooling Systems}
Dry air cooling systems reject waste heat by releasing it directly into the atmosphere without any water withdrawals or consumption. Given the lack of water withdrawal and consumption, the water footprint of dry cooling was set to zero in this study. Dry cooling systems require large heat exchangers making them significantly more expensive than recirculating cooling systems. Additionally, their efficiency depends greatly on ambient air temperatures and makes them less suitable during hot days when electricity demand is often at its highest.
\subsubsection{K: Evaporative Losses}
\paragraph{Once-Through System}
In once-through cooling systems, the fraction of water consumed downstream through evaporation, $k_{evap}$ is considered negligible. Consequently, the total water consumption for once-through cooling systems is set to zero.

\paragraph{Recirculating System}
Water consumed by recirculating cooling systems is expressed as follows\cite{Rutberg:2012:00,Rutberg:2011:00}:
\begin{align}\label{ewn:evaplosses}
& K_{k}(t) = Q_{Tk}(t)\bigg(\frac{1-k_{sens}}{h_{fg}}\bigg)\bigg(1+\frac{1-k_{bd}}{n_{cc}-1}\bigg)&
\end{align} 
where $k_{bd}$ represents the fraction of the blowdown that is treated and sent back to the water source. In this study, it is assumed that 100\% of the blowdown ($k_{bd}=1$) is treated and returned to the watershed. By substituting Equation \ref{ewm:ThermCool},  Equation \ref{ewn:evaplosses} becomes:
\begin{align}
& K_{k}(t) = J_{k}(t)\bigg(1-\eta_{k,other}\bigg)\bigg(\frac{1-k_{sens}}{h_{fg}}\bigg)&\label{ewn:evapfinal}
\end{align} 

\subsubsection{B: Non-Portable Water Withdrawal for Electrical Supply System}
Although it presents a significant opportunity for developing energy-water nexus synergies\cite{Siddiqi:2013:01,Siddiqi:2011:00}, this study assumes that none of the water withdrawals are from non-potable water sources.  

\section{A Case Study: The RTS-96 GMLC Test Case}\label{sec:RTS}
\begin{table}[!htb]
    \begin{minipage}{.43\linewidth}
        \centering
        \begin{tabular}{|lll|}\hline
        \textbf{Parameter}  & \textbf{Values} & \textbf{Units}          \\\hline
        $k_{os}$  others   & 20     & \%             \\
        $k_{os}$  nuclear   & 0     & \%             \\
        $k_{os}$  combined cycle   & 12     & \%             \\
        $n_{cc}$   & 6      & -              \\
        $c_{p,w}$  & 4.142  & MJ/kg$\cdot$ K \\
        $h_{fg}$   & 2.54   & MJ/kg          \\
        $\rho_w$   & 0.998  & $kg/m^3$       \\
        $\Delta_T$   & 10  & $^\circ$K       \\
        $k_{bd}$   & 0      & \%             \\
        $k_{evap}$ & 0      & \%           \\\hline 
        \end{tabular}
        \caption{Table of parameters values.\label{tb:1}}
    \end{minipage}%
    \begin{minipage}{.56\linewidth}
        \centering
        \begin{tabular}{|lll|}\hline
        \textbf{Resource Type}  & \textbf{Cost} & \textbf{Units}\\\hline
        Natural gas&3.8872&\$/MMBTU\\
        Oil&10.3494&\$/MMBTU\\
        Coal&2.1139&\$/MMBTU\\
        Nuclear&0.8104&\$/MMBTU\\
        Curtailable load&50&\$/MW\\
        Curtailable Hydro&2.5&\$/MW\\
        Curtailable Wind&0&\$/MW\\
        Curtailable Solar&1&\$/MW\\
        Active Demand Response&50&\$/MW\\
        Storage&0&\$/MWh\\\hline
        \end{tabular}
        \caption{Table of fuel, curtailment, active demand response and storage costs.\label{FuelCosts}}
    \end{minipage} 
\end{table}
The EPECS methodology summarized in Sections \ref{sec:scuc}-\ref{sec:water} has been tested and validated on slight modifications of the IEEE RTS-96 test case\cite{Muzhikyan:2015:SPG-J15,Muzhikyan:2015:SPG-J16} originally presented in \cite{Grigg:1999:00}. In this paper, a more recent version of the IEEE Reliability Test System (RTS-96) called the Reliability Test System Grid Modernization Laboratory Consortium (RTS-GMLC)\cite{DOE/NREL/-ALLIANCE:2019:00} is used to test and validate the methodology described above. Like the IEEE RTS-96 test case, the RTS-GMLC is comprised of 3 control areas, 73 buses, 99 generators with a maximum load capacity of 8550 MW. The new RTS-GMLC also includes wind, utility PV, rooftop PV, and hydro generation profiles. This test case also evolves the generation mix to reflect current grid generation mixes. For example, some of the oil and coal units are replaced with combined-cycle natural gas units both to minimize emissions and to support a high penetration of solar and wind.
\subsection{Overview}
As mentioned in the introduction, this paper seeks to understand the degree to which water infrastructure can provide flexibility to the electricity supply system.   The novelty of the EPECS methodology described above is in its accurate determination of operating reserves such as regulation, load-following and ramping reserves. Flexible control of water resources such as run-of-river hydro, conventional hydro, water and wastewater treatment facilities is specifically considered. For any given variable profile (e.g. hydro, solar, or wind), the ability to curtail the resource is also analyzed. Two scenarios are considered. As an ``experimental case", water and wastewater treatment facilities can provide demand response while run-of-river and conventional hydro resources are treated as curtailable resources.  That is, they provide load-following and ramping reserves through curtailment. In the ``control case", all water resources are considered inflexible. That is, run-of-river and conventional hydro are not curtailable, and water and wastewater treatment facilities cannot provide demand response. Based on the simulation results for the year, the amount of thermal generation is calculated and the resulting water withdrawals, and consumption is obtained. Additionally, the model can also estimate the amount of fuel used and subsequently the $CO_2$ emissions.
\subsubsection{Power and Water Resources}
The RTS-GMLC consists of 73 buses, 73 thermal generators, 20 hydro generating units, 56 solar units, 4 wind generators, 1 storage unit, supplying a peak load of 7979.5MW. Water resources considered in the study include all hydroelectric power plants as well as the electricity demand profile of water and wastewater treatment facilities. The load profile of water and wastewater treatment facilities was taken to be a fraction of the load profile eligible for curtailment and active demand response. The thermal generators were split into a set with once-through cooling and another with recirculating cooling systems. Table~\ref{tb:1} shows the assumed constants used in the calculation of water withdrawal, water consumption and total fuel consumed.
 \begin{figure}[!htb]
\centering
\includegraphics[width=6.2in] {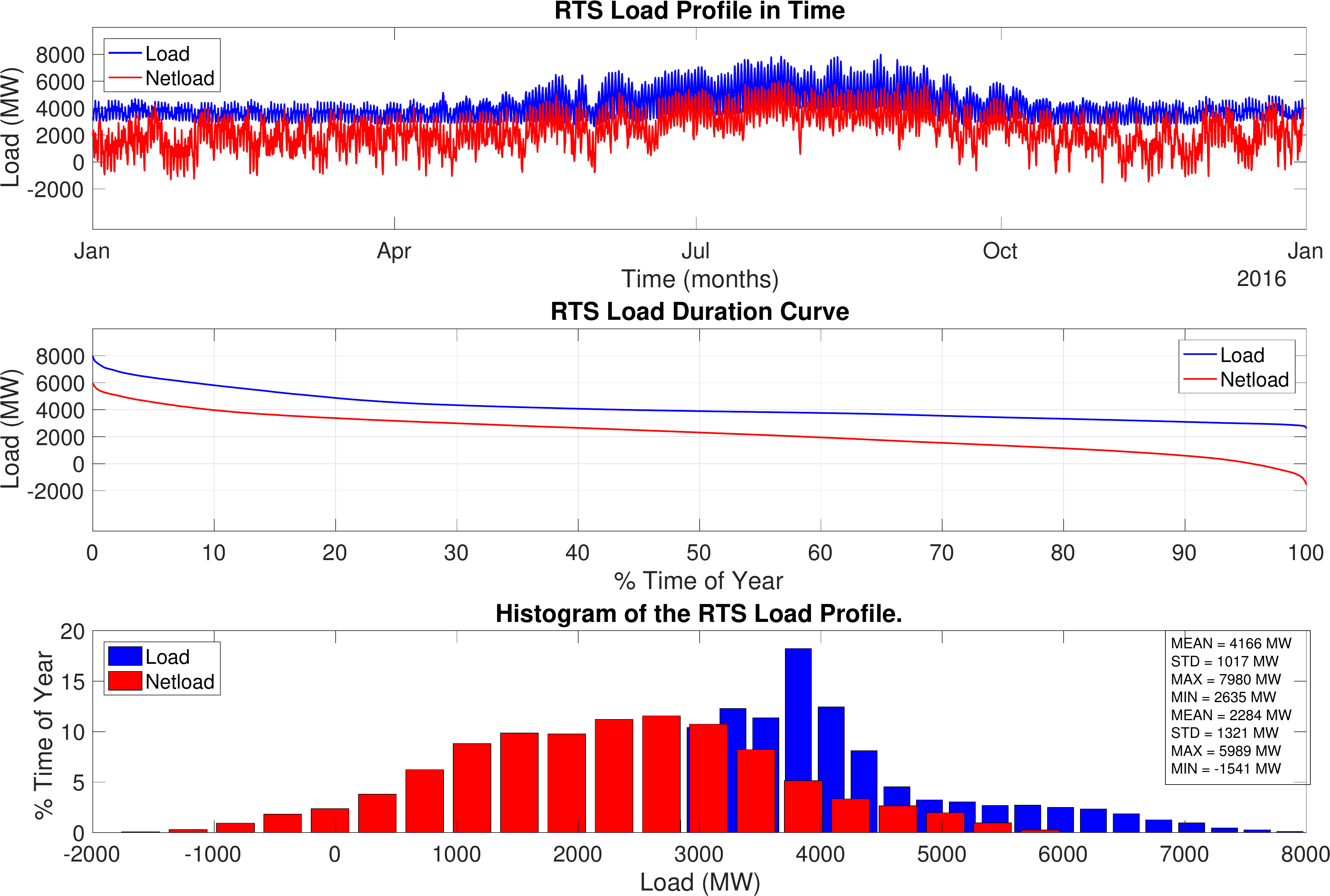}
\caption{Load and net Load profile for RTS-GMLC.} 
\label{fig:netload}
\end{figure}
\subsubsection{Heating Rate curves}
Heating curves for thermal generators are used to compute the fixed, linear and quadratic costs for generating electricity. These heating rate curves are later used to compute the fuel consumption and thermal load of thermal generating units. Table \ref{FuelCosts} provides the assumed fuel cost of all the resources used in this study.
\subsubsection{Time profiles}
Real-time and day-ahead time profiles for solar, wind, load and hydro generating facilities are also provided. These profiles are  used to compute the day-ahead and real-time forecasts which are then used as inputs to the optimization programs. As stated in Section \ref{sec:ver}, variable energy resources are analyzed based on the implementation introduced in \cite{Muzhikyan:2015:SPG-C46}. Given the actual renewable generation profile and expected errors, the day-ahead and real-time forecasts are computed.

Figure \ref{fig:netload} represents the net load distribution used in the RTS-GMLC test case. The first subplot represents the load profile in blue and the net load profile in red. Notice that in periods of low demand during the Spring and Fall months, the net load is very low and in some case less than zero MW. Negative net load represents cases when the generation exceeds the demand. Due to high amounts of variable renewable generation, the histogram of the net load in the third subplot is shifted further to the left and is negative for almost 40\% of the time as shown in the second subplot.
\section{Results}\label{sec:results}
\subsection{Load-Following Reserves}
\begin{figure}[!htb]
\centering
    \begin{minipage}{.495\textwidth}
        \centering
        \includegraphics[width=\textwidth] {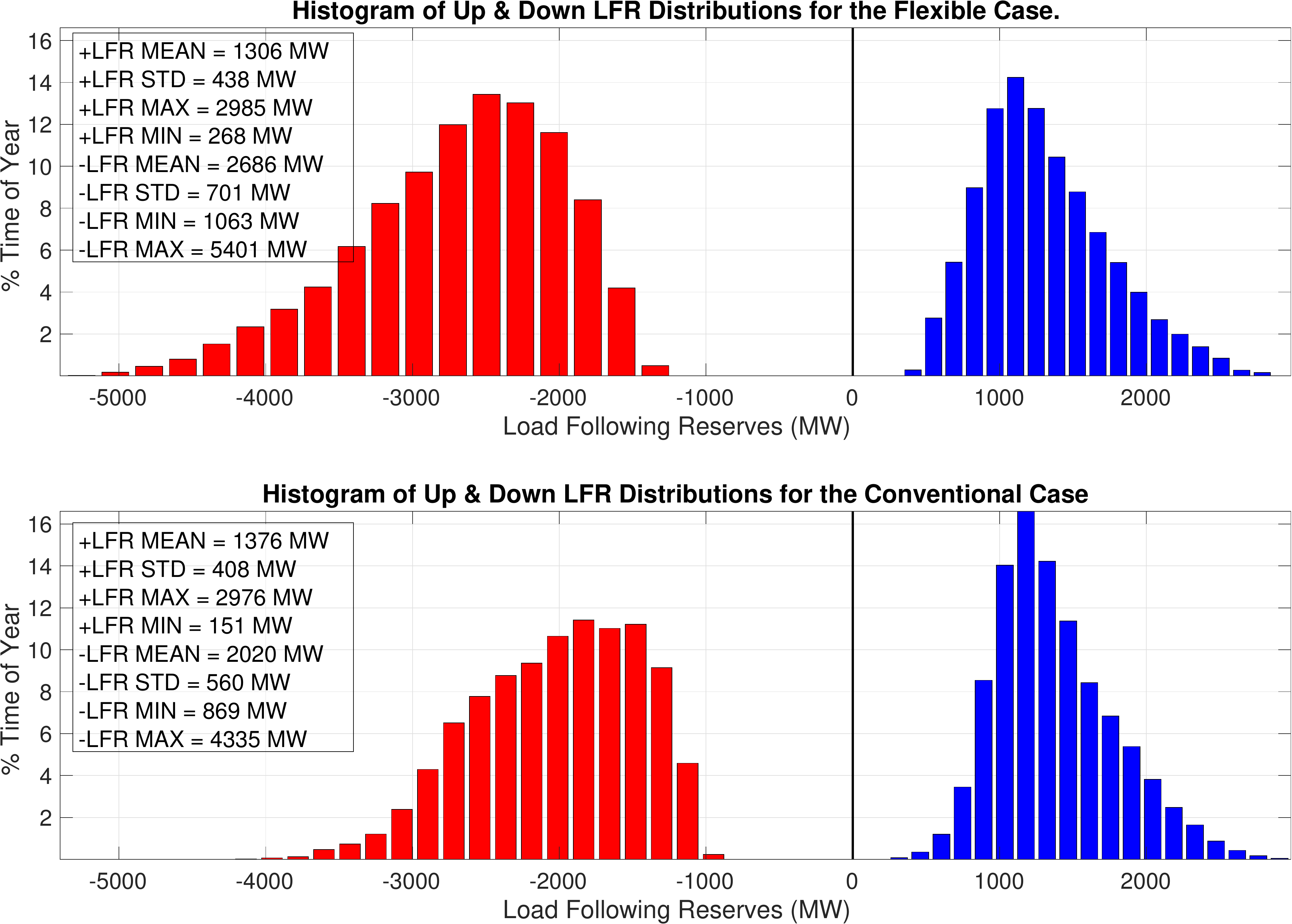}
        \caption{Histogram of load-following reserves for RTS-GMLC} 
        \label{fig:lfr}
    \end{minipage}%
    \begin{minipage}{.495\textwidth}
        \centering
        \includegraphics[width=\textwidth] {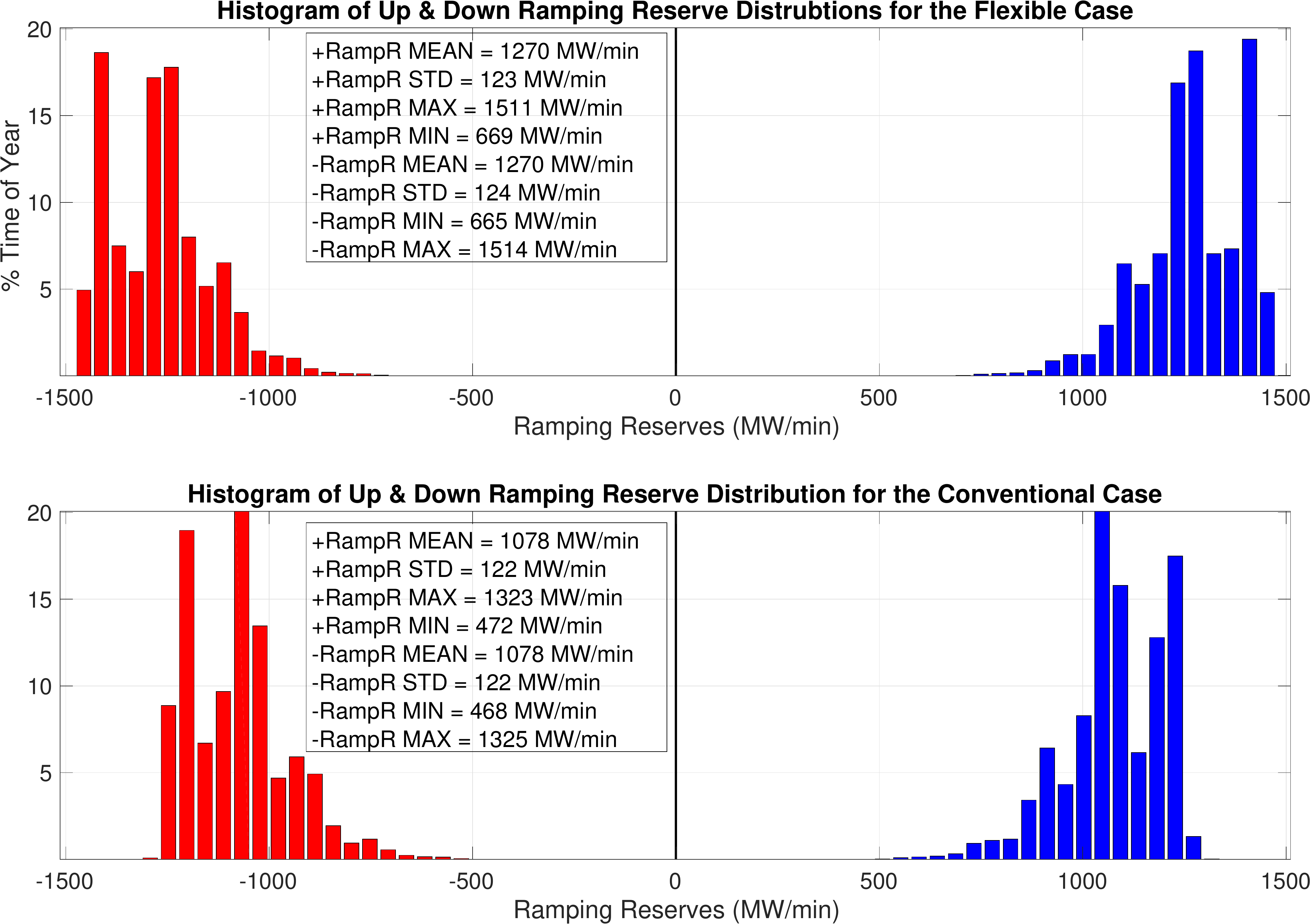}
        \caption{Histogram of ramping reserves for RTS-GMLC} 
        \label{fig:ramp}
    \end{minipage}
\end{figure}

Upward and downward load-following reserves are procured in the day-ahead market (SCUC). These reserves are then used in the real-time market to balance any variability in the net load. In this study, wind, solar and dispatchable generators contribute towards load-following reserves in the conventional case. While in the flexible case, run-of-river and conventional hydro also contribute towards load-following reserves through curtailment. Both downward and upward load-following reserves are equally important to ensure reliable operation of a system with a high penetration of variable renewable energy. Therefore, it is important that neither the upward nor downward load-following reserves are depleted. Figure~\ref{fig:lfr} shows a comparison of the load-following reserves profile for the conventional and flexible cases. Flexible operation of water resources increases the minimum levels of both the upward and download load following reserves so that the space between the red and blue distributions increases.  These larger minimum values of upward and downward load-following reserves improve system reliability because these reserves are not as close to being depleted.   

\subsection{Ramping Reserves}
Similar to the load-following reserves, upward and downward ramping reserves are also procured in the day-ahead market and used in the real-time market to balance variability in the net load. As before, wind, solar, dispatchable generators and hydro resources contribute towards ramping reserves in the flexible case. However, in the conventional case ramping reserves are not procured from hydro resources. As shown in Figure \ref{fig:ramp}, flexible operation improves the minimum levels of both upward and downward ramping reserves albeit by a small amount. This ensures that ramping reserves are not easily depleted in the presence of variable renewable generation.
\subsection{Regulation}
\begin{figure}[!htb]
\centering
    \begin{minipage}{.49\textwidth}
        \centering
        \includegraphics[width=\textwidth] {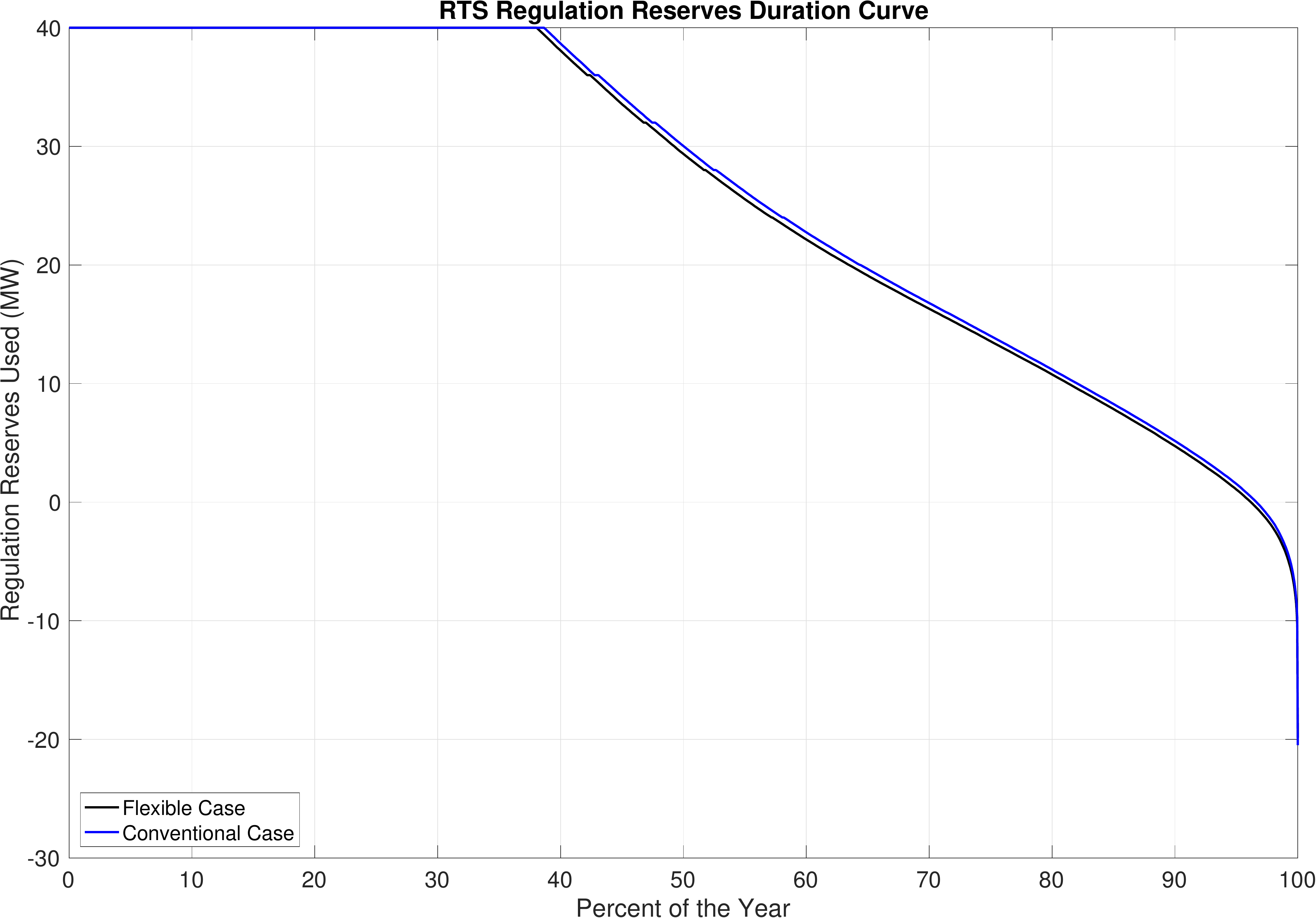}
        \caption{Regulation Reserves Duration Curve for RTS-GMLC} 
        \label{fig:reg}
    \end{minipage}%
    \begin{minipage}{.49\textwidth}
        \centering
        \includegraphics[width=\textwidth] {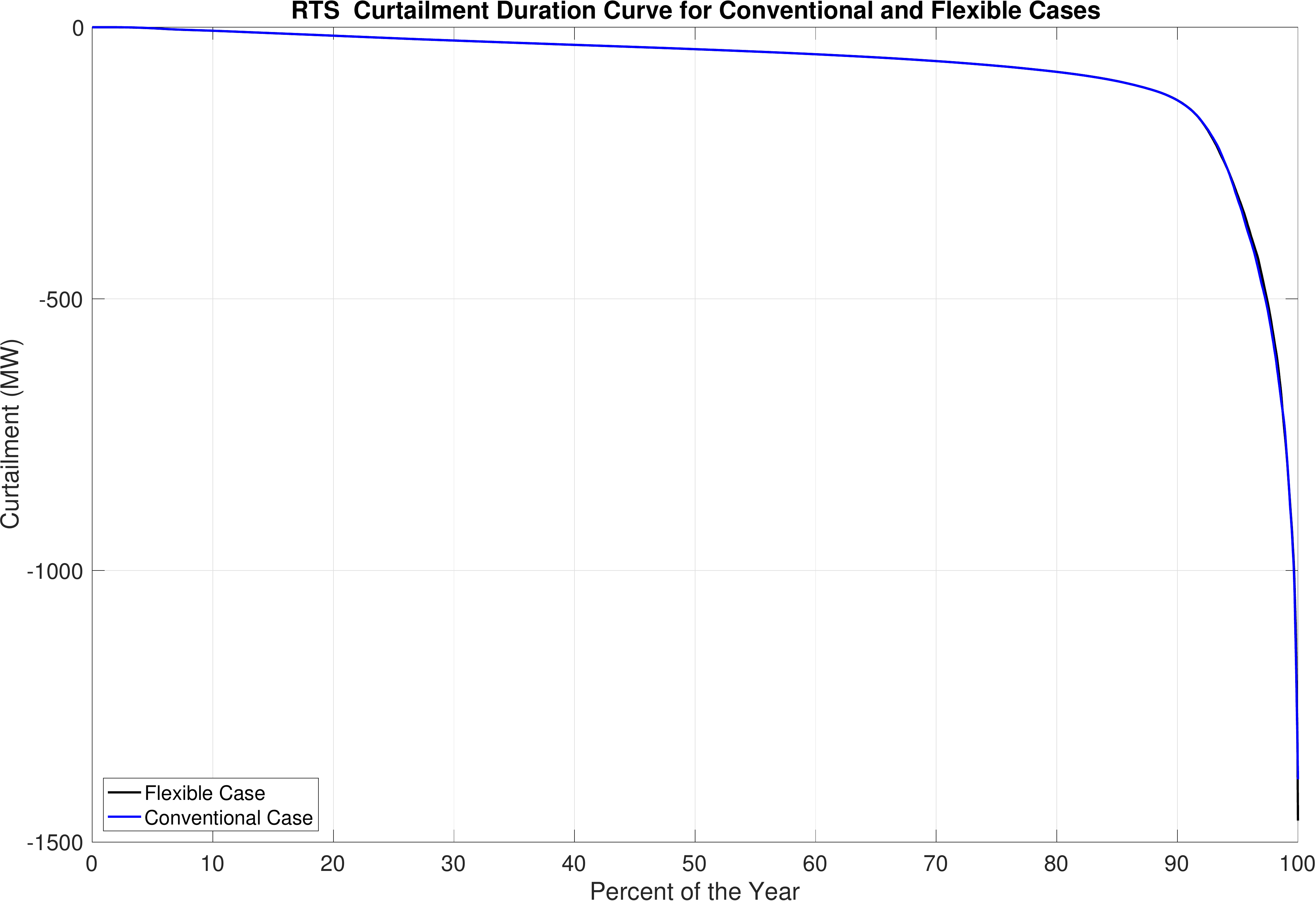}
        \caption{Curtailment Duration Curve for RTS-GMLC} 
        \label{fig:curt}
    \end{minipage}
\end{figure}

Regulation reserves are the fastest balancing resource and serve to mitigate system imbalances in real-time.  These reserves are used to balance the system after the application of load-following reserves, ramping reserves, curtailment have been utilized in the real-time market. As such, it is imperative that the system contains enough regulation reserves to mitigate imbalances. Figure~\ref{fig:reg} shows the regulation reserves as duration curves for both the conventional and flexible cases.  The system regulation capacity was set to $\pm$ 40MW.  As illustrated in Figure~\ref{fig:reg}, both the flexible and conventional cases show some saturation of regulation reserves in the upward direction. This indicates the need for more regulation reserves in the system. However, in the flexible case, the upward regulation is only saturated 38.4\% of the year whereas it is saturated 39.06\% of the year in the conventional case.  Interestingly, both the conventional and the flexible cases have no saturation of downward regulation for any time during the year. The difference in behavior between upward and downward regulation can largely be attributed to differences in the statistical characteristics of the net load time series.  
\subsection{Curtailment}
Curtailment of variable renewable generation serves a key balancing role especially in the absence of sufficient load-following and ramping reserves. In this study, water resources and hydro generation can be curtailed in the flexible case while in the conventional case, only solar and wind are curtailable.  Because the energy markets require balancing on a nodal basis, curtailment may also be caused by topological limitations of the power system.  In this study, flexible operation increases the overall curtailment amounts as now hydro resources are available for curtailment.  These new energy-water resources provide more system-wide flexibility as measured in terms of load-following and ramping reserves. The curtailment duration curves for both cases are shown in Figure~\ref{fig:curt}. Given the small amounts of hydro generation in the system, the overall curtailment in the two systems is rather similar on a total energy and percent-time basis as summarised in Table~\ref{tab:curtStats}.
\begin{table}[!t]
\caption{Curtailment statistics for the flexible and conventional cases.}
\begin{footnotesize}
\begin{center}
\begin{tabular}{p{2.7cm}rrrrr}\toprule
\textbf{Scenario}& \textbf{Tot. Res. (GWh)}& \textbf{Tot. Curt. Energy (GWh)}& \textbf{\% Energy Curt.}& \textbf{\% Time Curt.}& \textbf{Max Curt. Level (MW)}\\\toprule
\textbf{Flexible Case}& 16480 &   667 &  4.05 & 98.28 &  1460 \\\bottomrule
\textbf{Conventional Case}& 16480 &   672 &  4.08 & 97.94 &  1384 \\\bottomrule
\end{tabular}
\end{center}
\end{footnotesize}

\label{tab:curtStats}
\end{table}

\begin{figure}[!htb]
\centering
    \begin{minipage}{.49\textwidth}
        \centering
        \includegraphics[width=\textwidth] {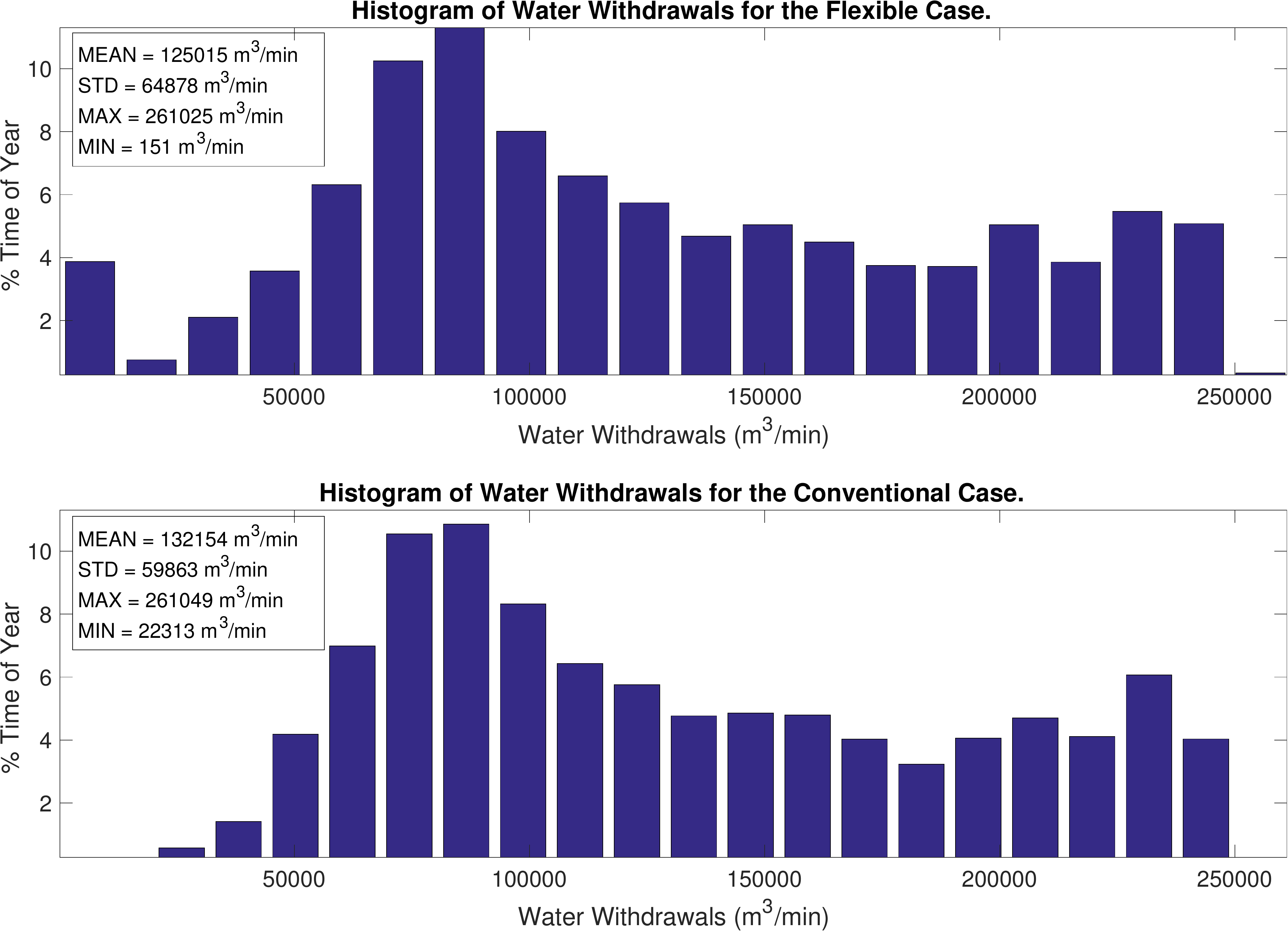}
        \caption{Histogram of water withdrawals.} 
        \label{fig:withdrawals}
    \end{minipage}
    \begin{minipage}{.49\textwidth}
        \centering
        \includegraphics[width=\textwidth] {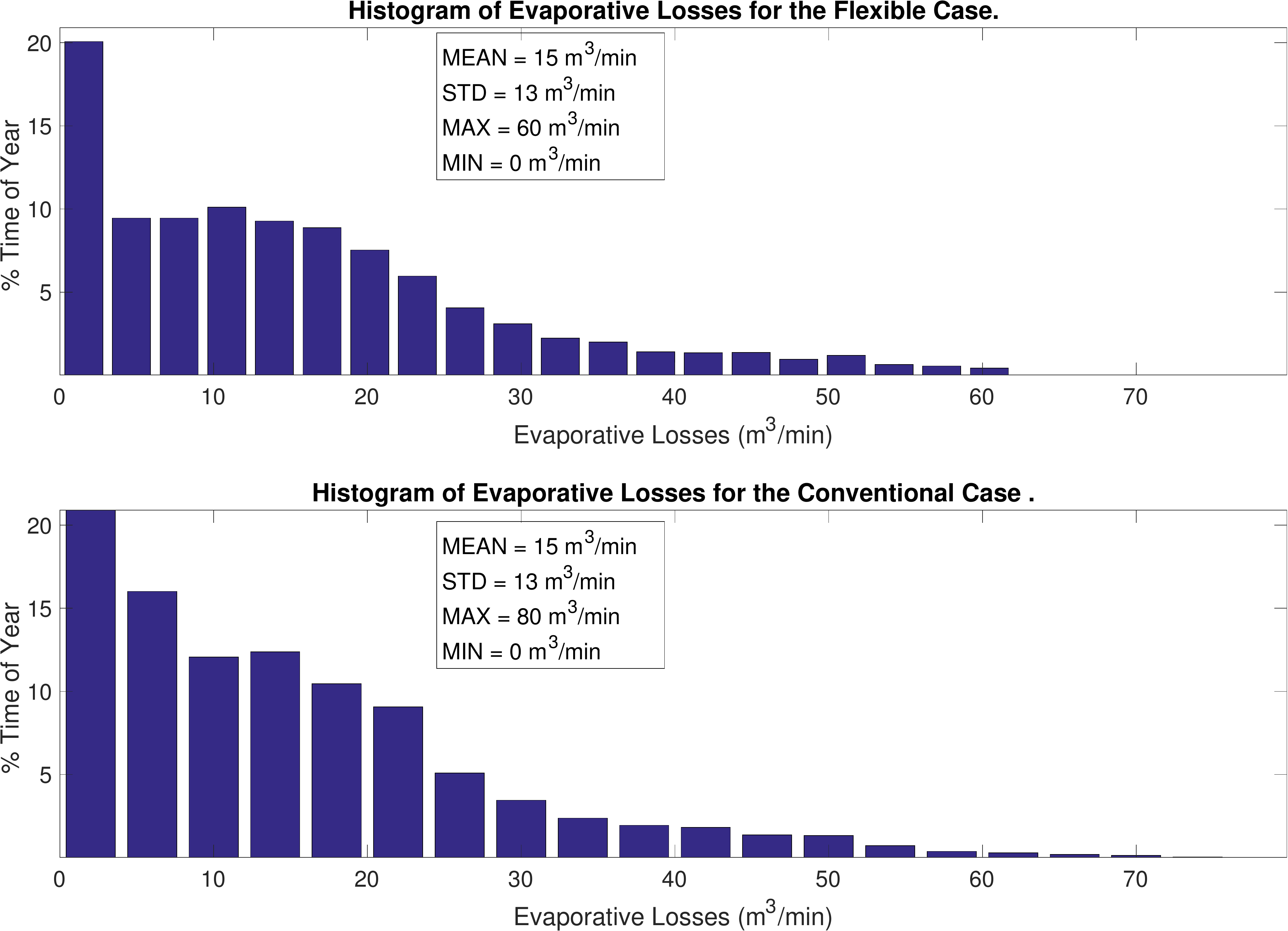}
        \caption{Histogram of evaporative losses.} 
        \label{fig:evap}
    \end{minipage}%
\end{figure}

\begin{figure}[!tbh]
\centering
\includegraphics[scale=.35] {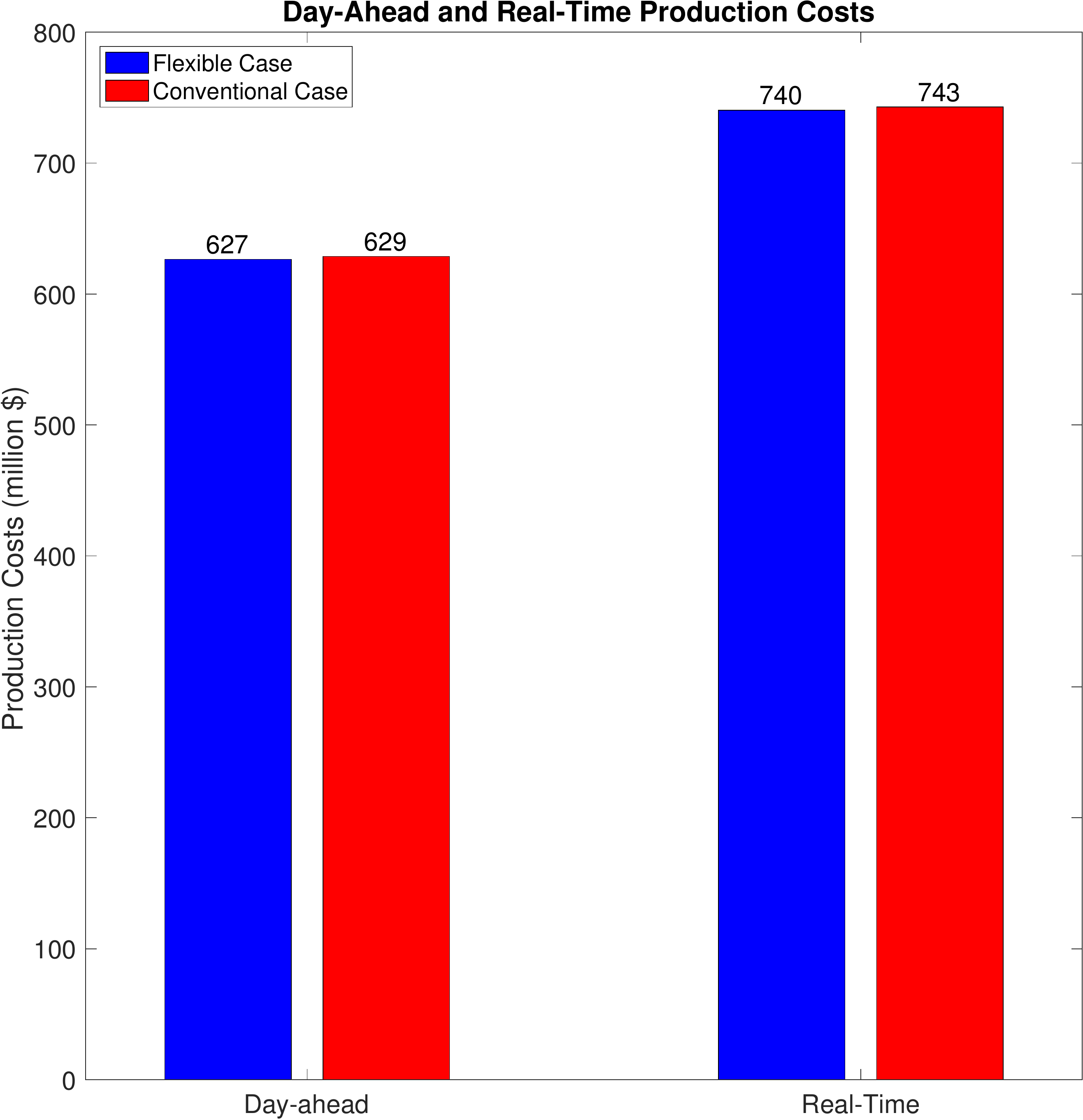}
\caption{Day-ahead and Real-time Production Costs.} 
\label{fig:prodCost}
\end{figure}
\subsection{Water Withdrawals and Consumption}
Water withdrawals and evaporative losses incurred by thermal power plants for cooling purposes are shown in Figures~\ref{fig:withdrawals} and \ref{fig:evap}. Flexible operation of water resources results in significantly lower amounts of water withdrawals and consumption.  On average, the conventional case withdraws 5.4\% more water than the flexible case.  Meanwhile, both cases have the same averages except the flexible case reduces the maximum water evaporation by 25\%.  The absolute values of water consumption are several orders of magnitude smaller than water consumption because of two factors.  First, the percentage of thermal power plants with recirculating cooling systems is small.  Second, the percentage of water evaporated is relatively small in comparison to the typical flow found in power plant cooling system.  

\subsection{Production Costs}
Flexible operation reduces the overall production cost of electricity in the both day-ahead market and in the real-time market although by only a few million dollars. Figure~\ref{fig:prodCost} compares the day-ahead and real-time production costs for both the conventional and flexible cases. As expected, the real-time production cost is slightly higher than the day-ahead production cost. However, in both markets the flexible case presents lower overall costs compared to the conventional case.

\subsection{Fuel Used}
Table~\ref{tab:fuelStats} represents the total fuel consumed in a year for each case including the percentage difference in fuel consumption. The fuel consumption results in Table~\ref{tab:fuelStats} illustrate that the overall fuel consumed in the flexible case is significantly lower by 2\% than in the conventional case. The flexible case utilizes natural gas units more than the conventional case while the conventional case uses a lot more coal (3\%) and oil (76\%).  Consequently, the conventional case generally has more units online than the flexible case.  
\begin{table}[!t]
\caption{Fuel statistics for the flexible and conventional cases.}
\begin{footnotesize}
\begin{center}
\begin{tabular}{rrrrrrr}\toprule
 & \textbf{Coal}& \textbf{NG}& \textbf{Nuclear}& \textbf{Oil}& \textbf{ Total }\\ \toprule
\textbf{Conventional Case} ($Kt^\pm$) &  4422 &  1123 &     0 &     4 &  5550 \\\midrule
\textbf{Flexible Case} ($Kt^\pm$) &  4306 &  1151 &     0 &     1 &  5458 \\\midrule
\textbf{Difference} ($Kt^\pm$) &   116 &   -28 &    -0 &     3 &    92 \\\midrule
\textbf{Percent Change} (\%)&     3 &    -2 &    -1 &    76 &     2 \\\bottomrule
\end{tabular}
\end{center}
\end{footnotesize}

\label{tab:fuelStats}
\end{table}
\subsection{Carbon Emissions}
Given its lower fuel consumption, the overall carbon emissions in the conventional case is significantly larger than in the flexible case. Flexible operation of hydro resources reduces the $CO_2$ emissions by 92 kilotons or 2\% as summarized in Table~\ref{tab:CO2Stats}.

\begin{table}[!t]
\caption{A comparison of $CO_2$ emissions for the flexible and conventional cases.}
\begin{footnotesize}
\begin{center}
\begin{tabular}{rrrrrrr}\toprule
 & \textbf{Coal}& \textbf{NG}& \textbf{Nuclear}& \textbf{Oil}& \textbf{ Total }\\ \toprule
\textbf{Conventional Case} ($Kt^\pm$) & 707420 & 180346 &     0 &   750 & 888516 \\\midrule
\textbf{Flexible Case} ($Kt^\pm$) & 688797 & 184775 &     0 &   184 & 873756 \\\midrule
\textbf{Difference} ($Kt^\pm$) & 18623 & -4429 &     0 &   567 & 14760 \\\midrule
\textbf{Percent Change} (\%)&     3 &    -2 &     0 &    76 &     2 \\\bottomrule
\end{tabular}
\end{center}
\end{footnotesize}

\label{tab:CO2Stats}
\end{table}
\section{Conclusion and Future Work}\label{sec:conc}
This study explored the degree to which the water supply infrastructure can provide flexibility to the electricity supply system. An enterprise control methodology was applied to study the balancing performance of two scenarios; an ``experimental case" with flexible operation of energy-water resources and a ``control case" without their flexible operation. The results obtained showed significant improvements in balancing performance, fuel consumption and $CO_2$ emissions in the ``experimental case" as compared to the ``control case". The ``experimental case" also shows significantly lower water withdrawals rates compared to the control. In conclusion, flexible operation of water resources significantly improves the performance of the system with high penetrations of variable renewable generation. 

While this paper serves to primarily demonstrate the methodology on a tractable test case, future work would seek to apply this methodology to a full scale case study.  From a methodological perspective, this work can be extended to investigate the role of non-potable water and desalination facilities. The work could also be extended to incorporate a model of the natural water system (e.g. hydrological river flows) which is particularly important in the face of climate change. Another extension could incorporate a model of the built water system so as to get a better understanding of the water utility operations and their overall effects on the nexus.

\section{Acknowledgements}
This paper was partially funded by the United States Department of Energy (US-DOE).
\bibliographystyle{IEEEtran} 
\bibliography{ewn,LIINESPublications,LIINESLibrary}

\begin{thebibliography}{100}
\providecommand{\url}[1]{#1}
\csname url@samestyle\endcsname
\providecommand{\newblock}{\relax}
\providecommand{\bibinfo}[2]{#2}
\providecommand{\BIBentrySTDinterwordspacing}{\spaceskip=0pt\relax}
\providecommand{\BIBentryALTinterwordstretchfactor}{4}
\providecommand{\BIBentryALTinterwordspacing}{\spaceskip=\fontdimen2\font plus
\BIBentryALTinterwordstretchfactor\fontdimen3\font minus
  \fontdimen4\font\relax}
\providecommand{\BIBforeignlanguage}[2]{{%
\expandafter\ifx\csname l@#1\endcsname\relax
\typeout{** WARNING: IEEEtran.bst: No hyphenation pattern has been}%
\typeout{** loaded for the language `#1'. Using the pattern for}%
\typeout{** the default language instead.}%
\else
\language=\csname l@#1\endcsname
\fi
#2}}
\providecommand{\BIBdecl}{\relax}
\BIBdecl

\bibitem{Rogers:2013:00}
J.~Rogers, K.~Averyt, S.~Clemmer, M.~Davis, F.~Flores-Lopez, D.~Kenney,
  J.~Macknick, N.~Madden, J.~Meldrum, S.~Sattler, and E.~Spanger-Siegfried,
  ``{Water-Smart Power: Strengthening the U.S. Electricity System in a Warming
  World},'' Union for Concerned Scientists, Cambridge, MA, Tech. Rep., 2013.

\bibitem{Averyt:2013:00}
K.~Averyt, J.~Macknick, J.~Rogers, N.~Madden, J.~Fisher, J.~Meldrum, and
  R.~Newmark, ``Water use for electricity in the united states: an analysis of
  reported and calculated water use information for 2008,'' \emph{Environmental
  Research Letters}, vol.~8, no.~1, p. 015001, Jan 2013.

\bibitem{Kanyerere:2018:00}
T.~Kanyerere, S.~Tramberend, A.~D. Levine, P.~Mokoena, P.~Mensah, W.~Chingombe,
  J.~Goldin, S.~Fatima, and M.~Prakash, ``Water futures and solutions: Options
  to enhance water security in sub-saharan africa,'' in \emph{Systems Analysis
  Approach for Complex Global Challenges}.\hskip 1em plus 0.5em minus
  0.4em\relax Springer, 2018, pp. 93--111.

\bibitem{Dodder:2016:00}
R.~S. Dodder, J.~T. Barnwell, and W.~H. Yelverton, ``Scenarios for low carbon
  and low water electric power plant operations: Implications for upstream
  water use,'' \emph{Environmental science \& technology}, vol.~50, no.~21, pp.
  11\,460--11\,470, 2016.

\bibitem{Al-Ansari:2017:00}
T.~Al-Ansari, A.~Korre, Z.~Nie, and N.~Shah, ``Integration of greenhouse gas
  control technologies within the energy, water and food nexus to enhance the
  environmental performance of food production systems,'' \emph{Journal of
  Cleaner Production}, 2017.

\bibitem{Al-Nory:2014:00}
M.~Al-Nory and M.~El-Beltagy, ``An energy management approach for renewable
  energy integration with power generation and water desalination,''
  \emph{Renewable Energy}, vol.~72, pp. 377--385, 2014.

\bibitem{Al-Aboosia:2018:00}
F.~Y. Al-Aboosia and M.~M. El-Halwagia, ``An integrated approach to
  water-energy nexus in shale-gas production,'' \emph{Processes}, vol.~6,
  p.~52, 2018.

\bibitem{Armstrong:2018:00}
N.~R. Armstrong, R.~C. Shallcross, K.~Ogden, S.~Snyder, A.~Achilli, and E.~L.
  Armstrong, ``Challenges and opportunities at the nexus of energy, water, and
  food: A perspective from the southwest united states,'' \emph{MRS Energy \&
  Sustainability}, vol.~5, 2018.

\bibitem{IEA:2017:01}
IEA, ``Key world energy statistics,'' {International Energy Agency}, Tech.
  Rep., September 2017.

\bibitem{IEA:2017:00}
------, ``Renewables 2017 analysis and forecasts to 2022,'' Tech. Rep., October
  2017.

\bibitem{Curtright:2008:01}
A.~E. Curtright and J.~Apt, ``The character of power output from utility-scale
  photovoltaic systems,'' \emph{Progress in Photovoltaics: Research and
  Applications}, vol.~16, no.~3, pp. 241--247, 2008.

\bibitem{Apt:2007:00}
J.~Apt and A.~Curtright, ``The spectrum of power from utility-scale wind farms
  and solar photovoltaic arrays,'' Carnegie Mellon Electricity Industry Center
  Working Paper, Pittsburgh, PA, United states, Tech. Rep., 2007.

\bibitem{Milano:2010:17}
\BIBentryALTinterwordspacing
F.~Milano, \emph{{Power system modelling and scripting}}, 1st~ed.\hskip 1em
  plus 0.5em minus 0.4em\relax New York: Springer, 2010. [Online]. Available:
  \url{http://www.uclm.es/area/gsee/web/Federico/psat.htm}
\BIBentrySTDinterwordspacing

\bibitem{Farid:2014:SPG-J26}
\BIBentryALTinterwordspacing
A.~M. Farid, B.~Jiang, A.~Muzhikyan, and K.~Youcef-Toumi, ``{The Need for
  Holistic Enterprise Control Assessment Methods for the Future Electricity
  Grid},'' \emph{Renewable \& Sustainable Energy Reviews}, vol.~56, no.~1, pp.
  669--685, 2015. [Online]. Available:
  \url{http://dx.doi.org/10.1016/j.rser.2015.11.007}
\BIBentrySTDinterwordspacing

\bibitem{Wolter:2012:00}
M.~Wolter, ``Reducing the curtailment amount of renewable energy using
  efficiency-based allocation,'' in \emph{Mathworks Energy and Utilities
  Virtual Conference}, 2012, pp. 1--2.

\bibitem{Wei:2016:00}
W.~Wei, N.~Li, J.~Wang, and S.~Mei, ``Estimating the probability of infeasible
  real-time dispatch without exact distributions of stochastic wind
  generations,'' \emph{IEEE Transactions on Power Systems}, vol.~31, no.~6, pp.
  5022--5032, 2016.

\bibitem{Muhanji:2018:SPG-J37}
\BIBentryALTinterwordspacing
S.~O. Muhanji, A.~Muzhikyan, and A.~M. Farid, ``{Distributed Control for
  Distributed Energy Resources: Long-Term Challenges \& Lessons Learned},''
  \emph{IEEE Access}, vol.~6, no.~1, pp. 32\,737 -- 32\,753, 2018. [Online].
  Available: \url{http://dx.doi.org/10.1109/ACCESS.2018.2843720}
\BIBentrySTDinterwordspacing

\bibitem{DeMeo:2005:00}
E.~A. DeMeo, W.~Grant, M.~R. Milligan, and M.~J. Schuerger, ``{Wind plant
  integration},'' \emph{Power and Energy Magazine, IEEE}, vol.~3, no.~6, pp.
  38--46, 2005.

\bibitem{Smith:2007:00}
J.~C. Smith, M.~R. Milligan, E.~A. DeMeo, and B.~Parsons, ``{Utility Wind
  Integration and Operating Impact State of the Art},'' \emph{Power Systems,
  IEEE Transactions on}, vol.~22, no.~3, pp. 900--908, 2007.

\bibitem{Bird:2012:00}
L.~Bird, M.~Milligan, and NREL, ``Lessons from large-scale renewable energy
  integration studies preprint,'' in \emph{2012 World Renewable Energy Forum},
  no. June, Denver, CO, United states, 2012, pp. ----8.

\bibitem{Brouwer:2014:00}
A.~S. Brouwer, M.~van~den Broek, A.~Seebregts, and A.~Faaij, ``{Impacts of
  large-scale Intermittent Renewable Energy Sources on electricity systems ,
  and how these can be modeled},'' \emph{Renewable and Sustainable Energy
  Reviews}, vol.~33, pp. 443--466, 2014.

\bibitem{Lubega:2016:00}
W.~N. Lubega and A.~M. Farid, ``A reference system architecture for the
  energy--water nexus,'' \emph{{IEEE Systems Journal}}, vol.~10, no.~1, pp.
  106--116, 2016.

\bibitem{IEA:2017:05}
\BIBentryALTinterwordspacing
IEA. (2017, January 13) Natural gas prices in 2016 were the lowest in nearly 20
  years. [Online]. Available:
  \url{https://www.eia.gov/todayinenergy/detail.php?id=29552}
\BIBentrySTDinterwordspacing

\bibitem{Dierauer:2018:00}
J.~R. Dierauer, D.~M. Allen, and P.~H. Whitfield, ``Exploring future water
  demand and climate change impacts on water availability in the peace region
  of british columbia, canada,'' in \emph{The Water-Energy-Food Nexus}.\hskip
  1em plus 0.5em minus 0.4em\relax Springer, 2018, pp. 45--54.

\bibitem{Al-Douri:2017:00}
\BIBentryALTinterwordspacing
A.~Al-Douri, D.~Sengupta, and M.~El-Halwagi, ``Shale gas monetization - a
  review of downstream processing to chemicals and fuels,'' \emph{Journal of
  Natural Gas Science and Engineering}, 2017. [Online]. Available:
  \url{http://www.sciencedirect.com/science/article/pii/S1875510017302299}
\BIBentrySTDinterwordspacing

\bibitem{Kiaghadi:2017:00}
\BIBentryALTinterwordspacing
A.~Kiaghadi, R.~S. Sobel, and H.~S. Rifai, ``Modeling geothermal energy
  efficiency from abandoned oil and gas wells to desalinate produced water,''
  \emph{Desalination}, vol. 414, pp. 51 -- 62, 2017. [Online]. Available:
  \url{http://www.sciencedirect.com/science/article/pii/S001191641631757X}
\BIBentrySTDinterwordspacing

\bibitem{Nichols:2017:00}
K.~Nichols, J.~Sawyer, J.~Bruening, B.~Halldorson, K.~Madhavan \emph{et~al.},
  ``{Development of a Large Scale Water Recycling Program for the Delaware
  Basin, New Mexico},'' in \emph{SPE Health, Safety, Security, Environment, \&
  Social Responsibility Conference-North America}.\hskip 1em plus 0.5em minus
  0.4em\relax Society of Petroleum Engineers, 2017, pp. 1--12.

\bibitem{Chang:2015:00}
\BIBentryALTinterwordspacing
Y.~Chang, R.~Huang, R.~J. Ries, and E.~Masanet, ``Life-cycle comparison of
  greenhouse gas emissions and water consumption for coal and shale gas fired
  power generation in {C}hina,'' \emph{Energy}, vol.~86, no.~0, pp. 335--343,
  2015. [Online]. Available:
  \url{http://www.sciencedirect.com/science/article/pii/S0360544215004880}
\BIBentrySTDinterwordspacing

\bibitem{Siddiqi:2013:01}
\BIBentryALTinterwordspacing
A.~Siddiqi, A.~Kajenthira, and L.~D. Anadon,
  ``\BIBforeignlanguage{English}{Bridging decision networks for integrated
  water and energy planning},'' \emph{\BIBforeignlanguage{English}{Energy
  Strategy Reviews}}, vol.~2, no.~1, pp. 46 -- 58, 2013, agricultural
  sector;Boundary spanning;Integrated resources;Inter-organizational
  network;International donors;Reuse of wastewater;Strategic
  objectives;Water-energy nexus;. [Online]. Available:
  \url{http://dx.doi.org/10.1016/j.esr.2013.02.003}
\BIBentrySTDinterwordspacing

\bibitem{Sanders:2014:00}
K.~T. Sanders, M.~F. Blackhurst, C.~W. King, and M.~E. Webber, ``The impact of
  water use fees on dispatching and water requirements for water-cooled power
  plants in texas,'' \emph{Environ. Sci. Technol.}, vol.~48, no.~12, p.
  140602120931006, Jun 2014.

\bibitem{Ela:2009:00}
E.~Ela, M.~Milligan, B.~Parsons, D.~Lew, and D.~Corbus, ``The evolution of wind
  power integration studies: past, present, and future,'' in \emph{Power \&
  Energy Society General Meeting, 2009. PES'09. IEEE}.\hskip 1em plus 0.5em
  minus 0.4em\relax IEEE, 2009, pp. 1--8.

\bibitem{Holttinen:2012:01}
H.~Holttinen, M.~O. Malley, J.~Dillon, and D.~Flynn, ``Recommendations for wind
  integration studies -- {IEA} task 25,'' International Energy Agency,
  Helsinki, Tech. Rep., 2012.

\bibitem{Holttinen:2013:00}
H.~Holttinen, A.~Orths, H.~Abilgaard, F.~van Hulle, J.~Kiviluoma, B.~Lange,
  M.~OMalley, D.~Flynn, A.~Keane, J.~Dillon, E.~M. Carlini, J.~O. Tande,
  A.~Estanquiro, E.~G. Lazaro, L.~Soder, M.~Milligan, C.~Smith, and C.~Clark,
  ``Iea wind export group report on recommended practices wind integration
  studies,'' International Energy Agency, Paris, France, Tech. Rep., 2013.

\bibitem{Muzhikyan:2019:SPG-JR04}
A.~Muzhikyan, S.~Muhanji, G.~Moynihan, D.~Thompson, Z.~Berzolla, and A.~M.
  Farid, ``{The 2017 ISO New England System Operational Analysis and Renewable
  Energy Integration Study},'' \emph{{Energy Reports}}, vol.~5, pp. 747--792,
  {July} 2019.

\bibitem{GE-Energy:2010:01}
GE-Energy, ``New england wind integration study,'' GE Energy and ISO New
  England, Schenectady, New York, Tech. Rep. May, 2010.

\bibitem{Shlatz:2011:00}
E.~Shlatz, L.~Frantzis, T.~McClive, G.~Karlson, D.~Acharya, S.~Lu, P.~Etingov,
  R.~Diao, J.~Ma, N.~Samaan, V.~Chadliev, M.~Smart, R.~Salgo, R.~Sorensen,
  B.~Allen, B.~Idelchik, A.~Ellis, J.~Stein, C.~Hanson, Y.~V. Makarov, X.~Guo,
  R.~P. Hafen, C.~Jin, and H.~Kirkham, ``Large-scale pv integration study,''
  Navigant Consulting, Las Vegas, NV, USA, Tech. Rep., 2011.

\bibitem{Piwko:2005:00}
R.~Piwko, X.~Bai, K.~Clark, G.~Jordan, N.~Miller, and J.~Zimberlin, ``{The
  Effects of Integrating Wind Power on Transmission System Planning,
  Reliability and Operations},'' GE Energy, Schnectady, New York, Tech. Rep.,
  2005.

\bibitem{Holttinen:2008:01}
H.~Holttinen, M.~Milligan, B.~Kirby, T.~Acker, V.~Neimane, and T.~Molinski,
  ``Using standard deviation as a measure of increased operational reserve
  requirement for wind power by wind engineering using standard deviation as a
  measure of increased operational reserve requirement for wind power,''
  vol.~44, no.~0, 2008.

\bibitem{Robitaille:2012:00}
A.~Robitaille, I.~Kamwa, A.~H. Oussedik, M.~de~Montigny, N.~Menemenlis,
  M.~Huneault, A.~Forcione, R.~Mailhot, J.~Bourret, and L.~Bernier,
  ``Preliminary impacts of wind power integration in the hydro-quebec system,''
  \emph{Wind Engineering}, vol.~36, no.~1, pp. 35--52, Feb 2012.

\bibitem{Ummels:2007:00}
\BIBentryALTinterwordspacing
B.~C. Ummels, M.~Gibescu, E.~Pelgrum, W.~L. Kling, and A.~J. Brand, ``{Impacts
  of Wind Power on Thermal Generation Unit Commitment and Dispatch},''
  \emph{IIEEE Transactions on Energy Conversion}, vol.~22, no.~1, pp. 44--51,
  2007. [Online]. Available: \url{http://dx.doi.org/10.1109/TEC.2006.889616
  http://ieeexplore.ieee.org/lpdocs/epic03/wrapper.htm?arnumber=4106021}
\BIBentrySTDinterwordspacing

\bibitem{Al-Saidi:2018:00}
M.~Al-Saidi and A.~Hefny, ``Institutional arrangements for beneficial regional
  cooperation on water, energy and food priority issues in the eastern nile
  basin,'' \emph{Journal of Hydrology}, 2018.

\bibitem{Burnett:2018:00}
K.~Burnett and C.~A. Wada, ``Accounting for externalities in the water energy
  food nexus,'' in \emph{The Water-Energy-Food Nexus}.\hskip 1em plus 0.5em
  minus 0.4em\relax Springer, 2018, pp. 261--272.

\bibitem{Eren:2018:00}
A.~Eren, ``Transformation of the water-energy nexus in turkey: re-imagining
  hydroelectricity infrastructure,'' \emph{Energy Research \& Social Science},
  2018.

\bibitem{Gurdak:2018:00}
J.~J. Gurdak, ``The water-energy-food nexus and california's sustainable
  groundwater management act,'' in \emph{The Water-Energy-Food Nexus}.\hskip
  1em plus 0.5em minus 0.4em\relax Springer, 2018, pp. 145--155.

\bibitem{Gleick:2017:00}
P.~Gleick, ``{Impacts of California's Five-Year (2012-2016) Drought on
  Hydroelectricity Generation},'' Pacific Institute, Oakland, California, Tech.
  Rep., 2017.

\bibitem{Pincetl:2015:00}
S.~Pincetl, R.~Graham, S.~Murphy, and D.~Sivaraman, ``Analysis of
  high-resolution utility data for understanding energy use in urban systems:
  The case of los angeles, california,'' \emph{Journal of Industrial Ecology},
  2015.

\bibitem{Escriva-Bou:2015:00}
\BIBentryALTinterwordspacing
A.~Escriva-Bou, J.~R. Lund, and M.~Pulido-Velazquez, ``Optimal residential
  water conservation strategies considering related energy in california,''
  \emph{Water Resources Research}, vol.~51, no.~6, pp. 4482--4498, 2015.
  [Online]. Available: \url{http://dx.doi.org/10.1002/2014WR016821}
\BIBentrySTDinterwordspacing

\bibitem{GEI-Consultants:2012:00}
{GEI Consultants}, ``California's water-energy nexus pathways to
  implementation,'' Water-Energy Team of the Governor's Climate Action Team,
  Tech. Rep., 2012.

\bibitem{Haynes:1989:01}
\BIBentryALTinterwordspacing
K.~E. Haynes and T.~D. Georgianna, ``{Risk assessment of water allocation and
  pollution treatment policies in a regional economy: reliability,
  vulnerability and resiliency in the Yellowstone Basin of Montana},''
  \emph{Computers, Environment and Urban Systems}, vol.~13, no.~2, pp. 75--94,
  1989. [Online]. Available:
  \url{http://dx.doi.org/10.1016/0198-9715(89)90036-7}
\BIBentrySTDinterwordspacing

\bibitem{Mahgary:1993:00}
\BIBentryALTinterwordspacing
Y.~E. Mahgary and E.~Tamminen, ``{Relevant methods for planning and
  optimization of power and water production systems of the Gulf countries BT -
  Proceedings of DESAL '92 Arabian Gulf Regional Water Desalination Symposium,
  November 15, 1992 - November 17, 1992},'' \emph{Desalination}, vol.~92, no.
  1-3, pp. 149--170, 1993. [Online]. Available:
  \url{http://dx.doi.org/10.1016/0011-9164(93)80079-3}
\BIBentrySTDinterwordspacing

\bibitem{Formiga:2003:00}
K.~T.~M. Formiga, F.~H. Chaudhry, P.~B. Cheung, and F.~R. Luisa, ``Optimal
  design of water distribution system by multiobjective evolutionary methods,''
  in \emph{Evolutionary Multi-Criterion Optimization}.\hskip 1em plus 0.5em
  minus 0.4em\relax Berlin, Germany: Springer Berlin Heidelberg, 2003, pp.
  677--691.

\bibitem{Diniz:2008:00}
A.~Diniz and M.~Maceira, ``A four-dimensional model of hydro generation for the
  short-term hydrothermal dispatch problem considering head and spillage
  effects,'' \emph{IEEE Trans. Power Syst.}, vol.~23, no.~3, pp. 1298--1308,
  Aug 2008.

\bibitem{Meldrum:2013:00}
J.~Meldrum, S.~Nettles-Anderson, G.~Heath, and J.~Macknick, ``Life cycle water
  use for electricity generation: a review and harmonization of literature
  estimates,'' \emph{Environmental Research Letters}, vol.~8, no.~1, p. 015031,
  Mar 2013.

\bibitem{Macknick:2012:00}
J.~Macknick, R.~Newmark, G.~Heath, and K.~C. Hallett, ``Operational water
  consumption and withdrawal factors for electricity generating technologies: a
  review of existing literature,'' \emph{Environmental Research Letters},
  vol.~7, no.~4, p. 045802, Dec 2012.

\bibitem{Macknick:2012:01}
J.~Macknick, S.~Sattler, K.~Averyt, S.~Clemmer, and J.~Rogers, ``The water
  implications of generating electricity: water use across the united states
  based on different electricity pathways through 2050,'' \emph{Environmental
  Research Letters}, vol.~7, no.~4, p. 045803, Dec 2012.

\bibitem{Averyt:2011:00}
K.~Averyt, J.~Fisher, A.~Huber-Lee, A.~Lewis, J.~Macknick, N.~Madden,
  J.~Rogers, and S.~Tellinghuisen, ``Freshwater use by us power plants:
  Electricity's thirst for a precious resource,'' Union of Concerned
  Scientists, Cambridge, MA, USA, Tech. Rep., 2011.

\bibitem{Bagloee:2018:00}
S.~A. Bagloee, M.~Asadi, and M.~Patriksson, ``Minimization of water pumps'
  electricity usage: a hybrid approach of regression models with
  optimization,'' \emph{Expert Systems with Applications}, 2018.

\bibitem{Bagirov:2013:00}
A.~M. Bagirov, A.~Barton, H.~Mala-Jetmarova, A.~Al~Nuaimat, S.~Ahmed,
  N.~Sultanova, and J.~Yearwood, ``An algorithm for minimization of pumping
  costs in water distribution systems using a novel approach to pump
  scheduling,'' \emph{Mathematical and Computer Modelling}, vol.~57, no. 3-4,
  pp. 873--886, 2013.

\bibitem{Ulanicki:2007:00}
B.~Ulanicki, J.~Kahler, and H.~See, ``Dynamic optimization approach for solving
  an optimal scheduling problem in water distribution systems,'' \emph{Journal
  of Water Resources Planning and Management}, vol. 133, no.~1, pp. 23--32,
  2007.

\bibitem{Lopez-Ibanez:2008:00}
M.~L{\'o}pez-Ib{\'a}{\~n}ez, T.~D. Prasad, and B.~Paechter, ``Ant colony
  optimization for optimal control of pumps in water distribution networks,''
  \emph{Journal of Water Resources Planning and Management}, vol. 134, no.~4,
  pp. 337--346, 2008.

\bibitem{Ghelichi:2018:00}
Z.~Ghelichi, J.~Tajik, and M.~S. Pishvaee, ``A novel robust optimization
  approach for an integrated municipal water distribution system design under
  uncertainty: A case study of mashhad,'' \emph{Computers \& Chemical
  Engineering}, vol. 110, pp. 13--34, 2018.

\bibitem{Menke:2016:00}
R.~Menke, E.~Abraham, P.~Parpas, and I.~Stoianov, ``Exploring optimal pump
  scheduling in water distribution networks with branch and bound methods,''
  \emph{Water Resources Management}, vol.~30, no.~14, pp. 5333--5349, 2016.

\bibitem{Lubega:2014:EWN-J11}
\BIBentryALTinterwordspacing
W.~N. Lubega and A.~M. Farid, ``{A Reference System Architecture for the
  Energy-Water Nexus},'' \emph{IEEE Systems Journal}, vol.~PP, no.~99, pp.
  1--11, 2014. [Online]. Available:
  \url{http://dx.doi.org/10.1109/JSYST.2014.2302031}
\BIBentrySTDinterwordspacing

\bibitem{Abdulla:2015:EWN-C53}
\BIBentryALTinterwordspacing
H.~Abdulla and A.~M. Farid, ``Extending the energy-water nexus reference
  architecture to the sustainable development of agriculture, industry \&
  commerce,'' in \emph{First IEEE International Smart Cities Conference},
  Guadalajara, Mexico, 2015, pp. 1--7. [Online]. Available:
  \url{http://dx.doi.org/10.1109/ISC2.2015.7366166}
\BIBentrySTDinterwordspacing

\bibitem{Lubega:2014:EWN-J12}
\BIBentryALTinterwordspacing
W.~N. Lubega and A.~M. Farid, ``{Quantitative Engineering Systems Model and
  Analysis of the Energy-Water Nexus},'' \emph{Applied Energy}, vol. 135,
  no.~1, pp. 142--157, 2014. [Online]. Available:
  \url{http://dx.doi.org/10.1016/j.apenergy.2014.07.101}
\BIBentrySTDinterwordspacing

\bibitem{Santhosh:2013:EWN-C16}
\BIBentryALTinterwordspacing
A.~Santhosh, A.~M. Farid, and K.~Youcef-Toumi, ``{Optimal Network Flow for the
  Supply Side of the Energy-Water Nexus},'' in \emph{2013 IEEE International
  Workshop on Intelligent Energy Systems}, Vienna, Austria, 2013, pp. 1--6.
  [Online]. Available:
  \url{http://dx.doi.org.libproxy.mit.edu/10.1109/IWIES.2013.6698578}
\BIBentrySTDinterwordspacing

\bibitem{Santhosh:2012:EWN-C09}
\BIBentryALTinterwordspacing
A.~Santhosh, A.~M. Farid, A.~Adegbege, and K.~Youcef-Toumi, ``{Simultaneous
  Co-optimization for the Economic Dispatch of Power and Water Networks},'' in
  \emph{The 9th IET International Conference on Advances in Power System
  Control, Operation and Management}, Hong Kong, China, 2012, pp. 1--6.
  [Online]. Available: \url{http://dx.doi.org/10.1049/cp.2012.2148}
\BIBentrySTDinterwordspacing

\bibitem{Hickman:2017:EWN-J32}
\BIBentryALTinterwordspacing
W.~Hickman, A.~Muzhikyan, and A.~M. Farid, ``{The Synergistic Role of Renewable
  Energy Integration into the Unit Commitment of the Energy Water Nexus},''
  \emph{Renewable Energy}, vol. 108, no.~1, pp. 220--229, 2017. [Online].
  Available: \url{https://dx.doi.org/10.1016/j.renene.2017.02.063}
\BIBentrySTDinterwordspacing

\bibitem{Diaz:2017:00}
C.~Diaz, F.~Ruiz, and D.~Patino, ``Modeling and control of water booster
  pressure systems as flexible loads for demand response,'' \emph{Applied
  Energy}, vol. 204, pp. 106--116, 2017.

\bibitem{Takahashi:2017:00}
S.~Takahashi, H.~Koibuchi, and S.~Adachi, ``Water supply operation and
  scheduling system with electric power demand response function,''
  \emph{Procedia Engineering}, vol. 186, pp. 327--332, 2017.

\bibitem{Menke:2017:00}
R.~Menke, E.~Abraham, P.~Parpas, and I.~Stoianov, ``Extending the envelope of
  demand response provision though variable speed pumps,'' \emph{Procedia
  Engineering}, vol. 186, pp. 584--591, 2017.

\bibitem{Menke:2016:01}
------, ``Demonstrating demand response from water distribution system through
  pump scheduling,'' \emph{Applied Energy}, vol. 170, pp. 377--387, 2016.

\bibitem{Muzhikyan:2015:SPG-C46}
\BIBentryALTinterwordspacing
A.~Muzhikyan, A.~M. Farid, and K.~Youcef-Toumi, ``{An Enhanced Method for
  Determination of the Ramping Reserves},'' in \emph{IEEE American Control
  Conference}, Los Angeles, CA, USA, 2015, pp. 1--8. [Online]. Available:
  \url{http://dx.doi.org/10.1109/ACC.2015.7170863}
\BIBentrySTDinterwordspacing

\bibitem{Muzhikyan:2015:SPG-C47}
\BIBentryALTinterwordspacing
------, ``{An Enhanced Method for Determination of the Regulation Reserves},''
  in \emph{IEEE American Control Conference}, Los Angeles, CA, USA, 2015, pp.
  1--8. [Online]. Available: \url{http://dx.doi.org/10.1109/ACC.2015.7170866}
\BIBentrySTDinterwordspacing

\bibitem{Muzhikyan:2015:SPG-J15}
\BIBentryALTinterwordspacing
------, ``{An Enterprise Control Assessment Method for Variable Energy Resource
  Induced Power System Imbalances Part 1: Methodology},'' \emph{IEEE
  Transactions on Industrial Electronics}, vol.~62, no.~4, pp. 2448--2458,
  2015. [Online]. Available: \url{http://dx.doi.org/10.1109/TIE.2015.2395391}
\BIBentrySTDinterwordspacing

\bibitem{Muzhikyan:2015:SPG-J16}
\BIBentryALTinterwordspacing
------, ``{An Enterprise Control Assessment Method for Variable Energy Resource
  Induced Power System Imbalances Part 2: Results},'' \emph{IEEE Transactions
  on Industrial Electronics}, vol.~62, no.~4, pp. 2459 -- 2467, 2015. [Online].
  Available: \url{http://dx.doi.org/10.1109/TIE.2015.2395380}
\BIBentrySTDinterwordspacing

\bibitem{Muzhikyan:2015:SPG-J22}
\BIBentryALTinterwordspacing
------, ``{Relative Merits of Load Following Reserves and Energy Storage Market
  Integration Towards Power System Imbalances},'' \emph{International Journal
  of Electrical Power and Energy Systems}, vol.~74, no.~1, pp. 222--229, 2016.
  [Online]. Available: \url{http://dx.doi.org/10.1016/j.ijepes.2015.07.013}
\BIBentrySTDinterwordspacing

\bibitem{Muzhikyan:2016:SPG-J28}
\BIBentryALTinterwordspacing
------, ``{An A Priori Analytical Method for Determination of Operating
  Reserves Requirements},'' \emph{International Journal of Energy and Power
  Systems}, vol.~86, no.~3, pp. 1--11, 2016. [Online]. Available:
  \url{http://dx.doi.org/10.1016/j.ijepes.2016.09.005}
\BIBentrySTDinterwordspacing

\bibitem{Farid:2017:SPG-IP46}
A.~M. Farid, ``2017 iso new england system operational analysis and renewable
  energy integration study (soares): Study methodology,'' in \emph{ISO New
  England Planning Advisory Committee}, Westborough, MA, 2017.

\bibitem{Farid:2017:SPG-IP47}
------, ``2017 iso new england system operational analysis and renewable energy
  integration study (soares): Scenario results,'' in \emph{ISO New England
  Planning Advisory Committee}, Westborough, MA, 2017.

\bibitem{Khawaji:2008:00}
A.~D. Khawaji, I.~K. Kutubkhanah, and J.-M. Wie, ``Advances in seawater
  desalination technologies,'' \emph{Desalination}, vol. 221, no. 1-3, pp.
  47--69, Mar 2008.

\bibitem{Al-Marafie:1989:00}
A.~Al-Marafie and M.~Darwish, ``Water production in kuwait --- its management
  and economics,'' \emph{Desalination}, vol.~71, no.~1, pp. 45--55, Jan 1989.

\bibitem{Shakouri:2010:00}
\BIBentryALTinterwordspacing
M.~Shakouri, H.~Ghadamian, and R.~Sheikholeslami, ``{Optimal model for multi
  effect desalination system integrated with gas turbine},''
  \emph{Desalination}, vol. 260, no. 1--3, pp. 254--263, 2010. [Online].
  Available:
  \url{http://www.sciencedirect.com/science/article/pii/S0011916410001670}
\BIBentrySTDinterwordspacing

\bibitem{Mezher:2011:01}
T.~Mezher, H.~Fath, Z.~Abbas, and A.~Khaled, ``Techno-economic assessment and
  environmental impacts of desalination technologies,'' \emph{Desalination},
  vol. 266, no. 1-3, pp. 263--273, Jan 2011.

\bibitem{Shakib:2012:00}
\BIBentryALTinterwordspacing
S.~E. Shakib, S.~R. Hosseini, M.~Amidpour, and C.~Aghanajafi,
  ``{Multi-objective optimization of a cogeneration plant for supplying given
  amount of power and fresh water},'' \emph{Desalination}, vol. 286, no.~0, pp.
  225--234, 2012. [Online]. Available:
  \url{http://www.sciencedirect.com/science/article/pii/S001191641100960X}
\BIBentrySTDinterwordspacing

\bibitem{Ghobeity:2014:00}
\BIBentryALTinterwordspacing
A.~Ghobeity and A.~Mitsos, ``Optimal design and operation of desalination
  systems: new challenges and recent advances,'' \emph{Current Opinion in
  Chemical Engineering}, vol.~6, no.~0, pp. 61 -- 68, 2014. [Online].
  Available:
  \url{http://www.sciencedirect.com/science/article/pii/S2211339814000793}
\BIBentrySTDinterwordspacing

\bibitem{Cardona:2002:00}
E.~Cardona, S.~Culotta, and A.~Piacentino, ``Energy saving with {MSF-RO} series
  desalination plants,'' \emph{Desalination}, vol. 153, pp. 167--171, 2002.

\bibitem{Pechtl:2003:00}
P.~Pechtl, M.~Dieleman, M.~Posch, B.~Davari, M.~Erbes, and S.~Schneeberger,
  ``Integrated thermal power and desalination plant optimization,'' in
  \emph{PowerGen Middle East}, Abu Dhabi, UAE, 2003, pp. 1--20.

\bibitem{Cardona:2004:00}
\BIBentryALTinterwordspacing
E.~Cardona and A.~Piacentino, ``{Optimal design of cogeneration plants for
  seawater desalination},'' \emph{Desalination}, vol. 166, no.~0, pp. 411--426,
  2004. [Online]. Available:
  \url{http://www.sciencedirect.com/science/article/pii/S0011916404003248}
\BIBentrySTDinterwordspacing

\bibitem{Sommariva:2007:00}
C.~Sommariva, ``{Thermal Desalination Processes and Economics},'' European
  Desalination Society, L'Aquila, Italy, Tech. Rep., 2007.

\bibitem{Gebel:2008:00}
J.~Gebel and S.~Y{\"u}ce, \emph{An Engineer's Guide to Desalination}.\hskip 1em
  plus 0.5em minus 0.4em\relax VGB PowerTech Service GmbH, 2008.

\bibitem{El-Nashar:2008:00}
A.~M. El-Nashar, ``Optimal design of a cogeneration plant for power and
  desalination taking equipment reliability into consideration,''
  \emph{Desalination}, vol. 229, no. 1-3, pp. 21--32, Sep 2008.

\bibitem{Rutberg:2011:00}
M.~J. Rutberg, A.~Delgado, H.~J. Herzog, and A.~F. Ghoniem, ``A system-level
  generic model of water use at power plants and its application to regional
  water use estimation,'' in \emph{ASME 2011 International Mechanical
  Engineering Congress and Exposition}.\hskip 1em plus 0.5em minus 0.4em\relax
  American Society of Mechanical Engineers, 2011, pp. 513--523.

\bibitem{Rutberg:2012:00}
M.~J. Rutberg, ``Modeling water use at thermoelectric power plants,'' Ph.D.
  dissertation, Massachussets Institute of Technology, 2012.

\bibitem{Integrated-Pollution-Prevention-and-Control-IPPC:2001:00}
{Integrated Pollution Prevention and Control (IPPC)} and E.~Commission,
  ``Reference document on the application of best available techniques to
  industrial cooling systems december 2001,'' European Commission, Tech. Rep.
  {December}, 2001.

\bibitem{ECOFYS:2014:00}
ECOFYS, ``-- geo-localised inventory of water use in cooling processes,
  assessment of vulnerability and of water use management measures,'' European
  Commission, End Report ENV.D1/SER/2013/0004, {July} 2014.

\bibitem{Eurelectric:1999:00}
Eurelectric, ``Bat for cooling systems, working group "environmental
  protection",'' Eurelectric, Tech. Rep., 1999.

\bibitem{Tsou:2013:00}
J.~L. Tsou, J.~Maulbetsch, J.~Shi, and EPRI, ``Power plant cooling system
  overview for researchers and technology developers,'' EPRI, Tech. Rep., 2013.

\bibitem{Pan:2018:01}
S.-Y. Pan, S.~W. Snyder, A.~I. Packman, Y.~J. Lin, and P.-C. Chiang, ``Cooling
  water use in thermoelectric power generation and its associated challenges
  for addressing water-energy nexus,'' \emph{Water-Energy Nexus}, vol.~1,
  no.~1, pp. 26--41, 2018.

\bibitem{King:2014:01}
C.~W. King, ``A systems approach for investigating water, energy, and food
  scenarios in east-central {Maui},'' Energy Institute, Jackson School of
  Geosciences, The University of Texas at Austin, Tech. Rep., 2014.

\bibitem{Siddiqi:2011:00}
A.~Siddiqi and L.~D. Anadon, ``The water--energy nexus in middle east and north
  africa,'' \emph{Energy Policy}, vol.~39, no.~8, pp. 4529--4540, Aug 2011.

\bibitem{Grigg:1999:00}
C.~Grigg, P.~Wong, P.~Albrecht, R.~Allan, M.~Bhavaraju, R.~Billinton, Q.~Chen,
  C.~Fong, S.~Haddad, S.~Kuruganty, W.~Li, R.~Mukerji, D.~Patton, N.~Rau,
  D.~Reppen, A.~Schneider, M.~Shahidehpour, and C.~Singh, ``{The IEEE
  Reliability Test System-1996. A report prepared by the Reliability Test
  System Task Force of the Application of Probability Methods Subcommittee},''
  \emph{Power Systems, IEEE Transactions on}, vol.~14, no.~3, pp. 1010--1020,
  1999.

\bibitem{DOE/NREL/-ALLIANCE:2019:00}
\BIBentryALTinterwordspacing
{DOE/NREL/ ALLIANCE}. (2019, April) {RTS-GMLC: Reliability Test System - Grid
  Modernization Lab Consortium}. [Online]. Available:
  \url{https://github.com/GridMod/RTS-GMLC}
\BIBentrySTDinterwordspacing

\end{thebibliography}
\end{document}